\documentclass[11pt]{article}
\usepackage[affil-it]{authblk}
\usepackage[margin=2cm]{geometry} 
\usepackage{lineno}
\usepackage{setspace}
\usepackage[utf8]{inputenc}
\usepackage[british]{babel} 
\usepackage[scaled]{helvet}
\usepackage[T1]{fontenc}
\usepackage{url}
\usepackage[colorlinks=true, allcolors=blue]{hyperref} 
\usepackage{amsmath,amsfonts,amssymb} 
\usepackage{siunitx}
\usepackage{graphicx}
\usepackage{caption}
\usepackage{multicol}
\usepackage{multirow}
\usepackage{threeparttable}
\usepackage[super,sort&compress,numbers]{natbib}
\usepackage{etoolbox}
\AtBeginEnvironment{thebibliography}{\linespread{1}\selectfont}

\begin{document}


\title{Accurate wave velocity measurement from diffuse wave fields}

\author{Melody Png$^{a*}$, Ming Huang$^a$, Marzieh Bahreman$^b$, Christopher M. Kube$^b$, Michael J. S. Lowe$^a$, Bo Lan$^{a*}$\footnote{\raggedright{\noindent Corresponding authors. E-mail addresses: m.png20@imperial.ac.uk (M. Png), bo.lan@imperial.ac.uk (B. Lan)}.}}
\affil{
$^a$ Department of Mechanical Engineering, Imperial College London, London SW7 2AZ, United Kingdom\\
$^b$ Department of Engineering Science and Mechanics, The Pennsylvania State University, University Park, 16802, Pennsylvania, USA\\
}
\date{\vspace{-5ex}}
\maketitle
       
\begin{abstract}	
\noindent Directional wave speeds variations in anisotropic elastic solids enables material characterisation capabilities, such as determination of elastic constants and volumetric measurement of crystallographic texture. However, achieving such measurements is challenging especially on samples with complex geometries. Here we propose the use of Green's Function reconstruction from diffuse ultrasonic wave fields for accurate velocity measurements on components with arbitrary geometries. Strategies for accurate reconstruction, including averaging over an increased number of different source locations, using longer window lengths of diffuse fields, and accurately deconvolving a source-dependent factor, were implemented to achieve satisfactory convergence towards Green's Function. Additionally, low signal intensity challenges from laser interferometers were overcome to enable non-contact measurement of the wave speeds, by making use of simultaneous excitation of sources to increase signal-to-noise ratio and signal normalisation to account for energy dissipation of diffuse fields. With successful demonstration using both phased array and laser receivers, this advancement fundamentally broadens acoustic wave velocity measurement capabilities to a wider range of environments and holds promise for future material characterisation of complex-shaped components.
\end{abstract}

\noindent \textbf{Keywords:} Ultrasound; Diffuse fields; Green's Function; Velocity measurement

\section{Introduction}
Elastic wave speeds in solids are an important material property with wide applications. For instance, measuring ultrasonic phase or group velocity in different directions through the inspected material enables the determination of elastic constants. This is crucial for designing mechanically-loaded components \cite{Ashby1999MaterialsDesign} and for monitoring the development of novel materials with enhanced mechanical characteristics, such as composites, and their utilisation in industrial applications \cite{Markham1970MeasurementUltrasonics, Smith1972UltrasonicComposites, Balasubramaniam1996UltrasonicComposites,  Rokhlinand1992DoubleMaterials}. Recent advancements have demonstrated that bulk texture information in polycrystalline hexagonal and cubic metals can also be inversely extracted from measured wave speeds in various directions throughout the material volume \cite{Lan2018DirectWaves}. These valuable applications highlight the importance and necessity of accurately measuring elastic wave speeds. 

Achieving accurate velocity measurements across different directions presents challenges, as it requires accurate knowledge of the distance and arrival time \cite{Rioux2018HighTechniques, Zou2018High-accuracyMonitoring}. This task is often made difficult when contact measurements are conducted with liquid couplant, which is detrimental to accurate measurement of Time-Of-Flight (TOF) at oblique angles, or when dealing with complex sample geometries, or both. The easiest and most typical method is to measure between two parallel flat surfaces, either using an immersion setup \cite{Hsu1992SimultaneousComposites, He2000MeasurementPulses, Carlson2003AnCement, Carreon2016RelationJoints} or a contact setup that is defined as a standard practice by the American Society for Testing and Materials (ASTM International) \cite{ASTM2022NondestructiveE2373}. This method provides highly accurate velocity measurements by leveraging the TOF between two successive backwall echoes, which heavily relies on the acquisition of clean multiple echoes. Consequently, it is only capable of measuring in the normal incidence direction or within a small angular range near it. To achieve accurate velocity measurements over a wide range of directions, a through-transmission and an improved double through-transmission configuration using an immersion setup have been developed. These setups reduce the effect of thickness variations from the use of couplant but are only effective on samples with parallel, flat surfaces \cite{Markham1970MeasurementUltrasonics, Rokhlinand1992DoubleMaterials}. The challenge posed by material geometry is evident in the double through-transmission method's sensitivity to sample surface parallelism and curvature, as discussed in a follow-up study \cite{Chu1994ComparativeMaterials}. Alternatively, a laser ultrasonic technique has been developed that uses lasers to generate and detect ultrasound at a distance. This enables non-contact determination of elastic constants and allows measurements to be conducted in high-temperature environments \cite{Aussel1989PrecisionDetermination, Audoin1997MeasurementTemperatures}. The use of lasers also facilitates the inspection of geometrically complex components \cite{Yawn2014ImprovedIndustry}. This will either require high-power laser sources to ablate the material and create a wave source, which is a destructive process; or lower-power lasers operating in the thermoelastic regime, which has a less favourable directivity pattern \cite{Krylov2016DirectivityParameters} that restricts the measurement of wave speeds to a narrow angular range.

In this paper, we propose the use of Green's Function (GF) reconstruction from diffuse ultrasonic fields, following the work by Lobkis and Weaver \cite{Lobkis2001OnField, Weaver2002OnPhonons}, for accurate velocity measurement on samples with arbitrarily shaped geometries. Diffuse fields in an enclosed structure can be described as wave fields that lost their coherence after multiple scattering from macroscopic sample geometries and microscopic grain boundaries. The amplitude and directional distribution of the waves are theoretically random and the waves are uncorrelated with respect to phase \cite{Evans1999MeasurementStructures}. The groundbreaking work in \cite{Lobkis2001OnField, Weaver2002OnPhonons} demonstrated that the ensemble average of the cross-correlation of responses at two specific locations within diffuse fields can effectively reconstruct the heterogeneous GF. This GF describes the relationship between the two specified points even though the wave fields recorded at these locations could be generated by either deterministic sources or random noise, such as thermal fluctuations. This makes it possible to carry out wave propagation evaluations even without an active source, which opened up a new paradigm of applications. 

The GF reconstruction method has enjoyed great success in seismology to study Earth's crust structures \cite{Gudmundsson2007Rayleigh-waveNoise, Behura2013VirtualInterferometry, Tanimoto2015InteractionIonosphere}. This stems from its advantage of not requiring a coherent wave source, effectively addressing the challenge of generating coherent fields for imaging in such environment, which heavily relies on explosions and natural earthquakes \cite{Campillo2003Long-RangeCoda, Shapiro2005High-ResolutionNoise}. This method has also been adopted in Non-Destructive Evaluation (NDE) applications, such as defect imaging \cite{Larose2006PassiveScatterer, Chehami2014DetectionCorrelation} and overcoming early-time saturation effects that obscure near-surface defects \cite{Potter2018DiffuseImaging, Zhang2019WavenumberFields, Zhang2019PhaseFields, Zhang2019Sparse-TFMStructures}. Due to its flexibility in terms of source excitation and sample geometry, it opens up the possibility of accurately measuring velocity in such settings, which is pursued in this paper.

The successes of this method in NDE application \cite{Larose2006PassiveScatterer, Zhang2019WavenumberFields} so far have mostly relied on amplitude summation but accurate velocity measurement poses much more stringent requirements on arrival times and is no trivial task to achieve. As will be elucidated in the next section, standard GF reconstruction can readily achieve defect imaging, but the resultant wave speed measurements still yield significant errors. To achieve error levels comparable to immersion ultrasonic measurements, we conducted an in-depth analysis to establish the necessary steps and techniques for achieving accurate reconstruction. Subsequently, we validated the proposed methodology with two independent setups: contact Phased Array (PA) receivers on flat surfaces and non-contact laser receivers on curved surfaces.

The paper is organised as follows: in Section 2, we introduce the methodology, discussing parameter optimisation, deconvolution of a source-dependent factor, and the application of signal processing techniques to enhance reconstruction accuracy. Additionally, we delve into the type of velocity measured through the GF reconstruction method. Section 3 provides a comprehensive description of the experimental procedures, covering both contact and non-contact receivers, and includes the presentation and discussion of velocity measurement results. Lastly, we explore the significance and potential implications of applying this methodology for velocity measurement before drawing conclusions.

\section{Methodology - GF Reconstruction from diffuse fields} \label{Section: Diffuse fields and GF reconstruction}

\subsection{GF Reconstruction theories} \label{Section: GF Reconstruction theories}

\begin{figure}[!t]
    \includegraphics[width=1\textwidth]{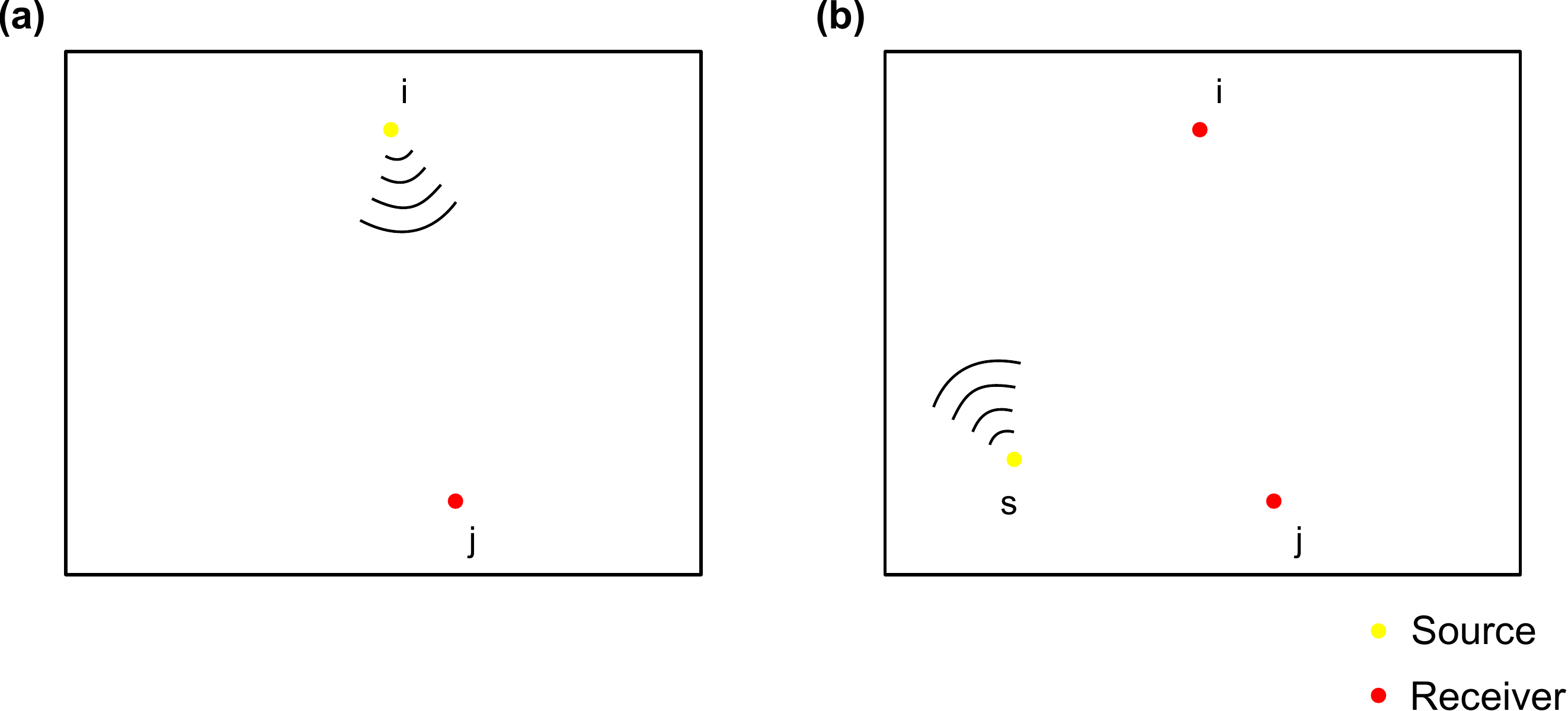}	
    \caption{Schematic illustrating the setups used for coherent and diffuse field measurements. \textbf{a} illustrates the setup used for typical coherent field measurements, where the coherent response is captured at location j from a source excitation at location i. \textbf{b} illustrates the setup used for diffuse field measurements, where diffused fields at locations i and j are captured from a source excitation at location s.}
    \label{fig:GF Reconstruction introduction}
\end{figure}
 
For conventional coherent field measurements, obtaining the coherent response typically involves transmitting a transient pulse from one location and receiving it at another. Referring to Figure \ref{fig:GF Reconstruction introduction}, this meant sending a pulse at location i and detecting it at location j. This coherent response can be reconstructed by cross-correlating the time-domain responses at locations i and j, obtained commonly from piezoelectric materials or laser interferometers in ultrasonic NDE, averaged over different locations of s. In the frequency domain, it may be similarly represented as shown in \cite{Lobkis2001OnField}.

\begin{equation}\label{eq:1}
    C_{j,i}\left(\omega\right)\approx\ \frac{R_{j, i}\left(\omega\right)}{F(-\omega)} = \frac{1}{N}\sum_{s=1}^{N}\frac{D_{i, s}(\omega) D_{j, s}^*(\omega)}{F(-\omega)}
\end{equation}

The early time coherent signal is $C_{j,i}(\omega)$ which is received at location $j$ and from an excitation at $i$, and $\omega$ represents the angular frequency. $R_{j, i}(\omega)$ is the diffuse field correlation, which can be calculated by multiplying a diffuse field term $D_{i, s}(\omega)$, with the conjugate of another $D_{j, s}^*(\omega)$. Here, $D_{i, s}(\omega)$ and $D_{j, s}(\omega)$ are the diffuse field data collected at $i$ and $j$, respectively. The approximation of $C_{j,i}(\omega)$ is obtained through the division of $R_{j, i}(\omega)$ by a source-dependent factor, $F(-\omega)$. $s$ denotes the indices of the source elements used to generate diffuse fields and $N$ is the total number of source elements used. The diffuse fields data  $D_{i, s}(\omega)$ and $D_{j, s}(\omega)$ are collected after some transmission delay $t_r$ and within a window length $T$. 

As demonstrated in \cite{Lobkis2001OnField} and evident in Equation \ref{eq:1}, the diffuse field correlation contains $F(-\omega)$, a source-dependent factor that needs to be deconvolved to achieve an accurate reconstruction. When a deterministic transient pulse is used as an excitation source, the frequency spectrum of the resultant diffuse field will be limited by the frequency characteristic of the pulse, instead of it being a flat spectral density over a wide range of frequencies. Consequently, the diffuse field correlation will inherently contain a transmission component that is related to the excitation source. Depending on the reconstruction aim—in this study, the GF convolved with the excitation pulse—the relevant source-dependent factor needs to be deconvolved to ensure high-accuracy reconstruction. Details about the source-dependent factor and the deconvolution process will be discussed further in Section \ref{Section: Deconvolution of source-dependant factor} and Section \ref{Section: Experimental implementation of GF reconstruction}.

\begin{figure}[!t]
  \centering
  \includegraphics[width=0.85\textwidth]{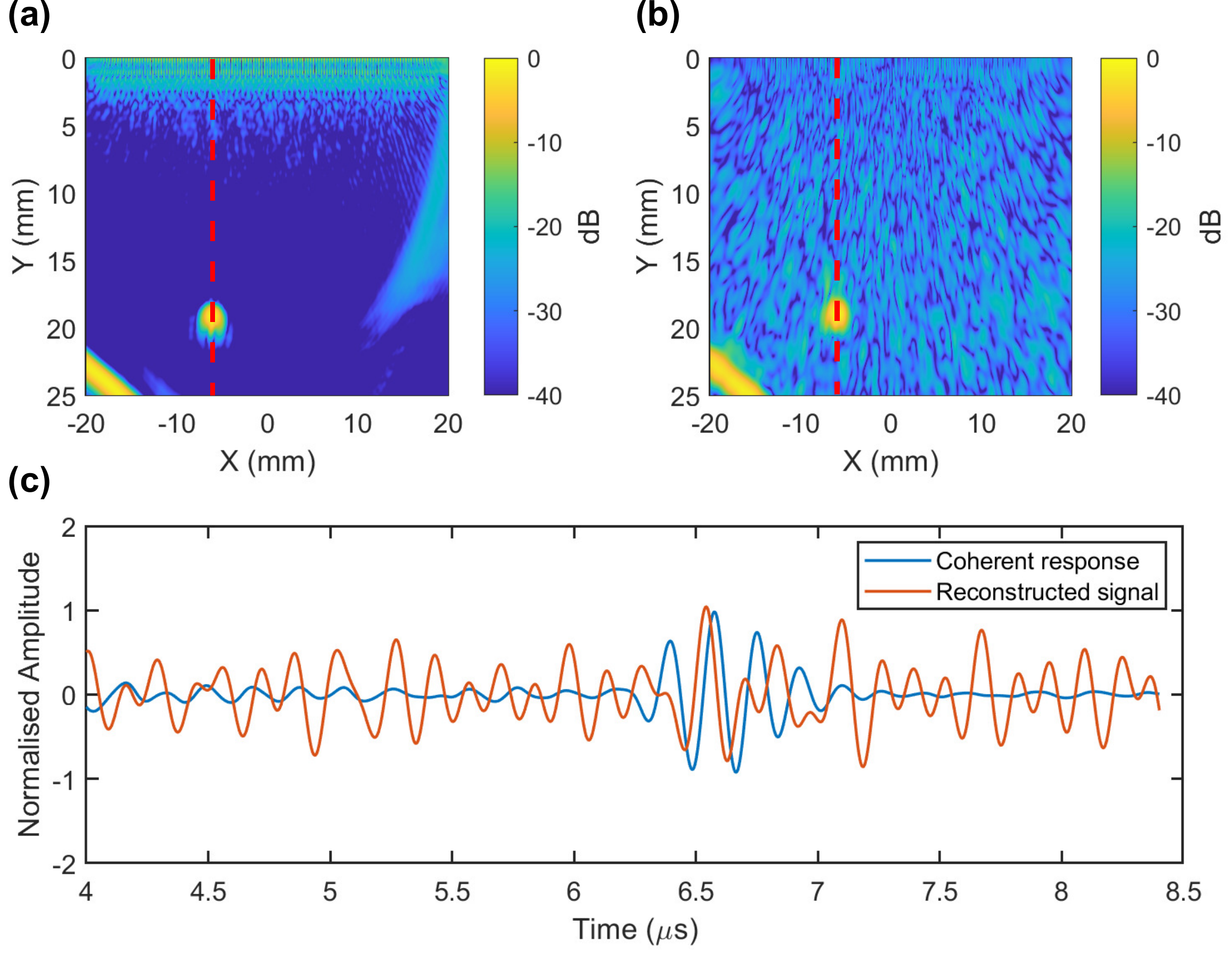}	
  \caption{Comparison of TFM images and time-domain signals obtained from both coherent and GF reconstruction method. \textbf{a} depicts a TFM image generated from coherent signals, revealing a defect at a depth of 19 mm. \textbf{b} displays the corresponding TFM image generated using the GF reconstruction method. \textbf{c} compares the pulse-echo responses obtained by one element of the array from both methods, where the wave pulse at 6.5 $\mu s$ represents the reflection from the defect. The red dashed lines indicate the x-coordinate where the pulse-echo response is obtained.}
  \label{fig: Reconstruction of TFM vs signal}
\end{figure}

Achieving high reconstruction accuracy is crucial for accurate velocity measurements, as deviations in waveform or a poor Signal-to-Noise Ratio (SNR) can impede the extraction of accurate arrival times essential for velocity calculations. This challenge is particularly pronounced when compared to the reconstruction of Total Focusing Method (TFM) images. For TFM, each point on the image is generated by summing all the data from each transmitter/receiver combination, acting as an averaging method to enhance the SNR of the image.  This property allows the GF reconstruction method to still be effective for defect detection using TFM images, even with a set of unoptimised parameters, as depicted in Figure \ref{fig: Reconstruction of TFM vs signal}a and b. The former illustrates a TFM image generated using coherent signals, revealing the presence of a defect at a depth of 19 mm. The latter shows the corresponding image generated using the GF reconstruction method. Despite the poorer SNR, the defect in the reconstructed image can still be identified due to this averaging principle. 
 
However, this principle cannot be leveraged when reconstructing individual signals. Upon examining the pulse-echo response at the location where the reflection from the defect is expected, as seen in Figure \ref{fig: Reconstruction of TFM vs signal}c, and comparing the two signals, it becomes apparent that the reflection signal at 6.5 \textmu s from the defect cannot be distinguished from the reconstruction noise. Essentially, this indicates that in the context of velocity measurement, accurate measurements using the TOF of transient pulses would not be achievable using the current approach. Therefore, it becomes crucial to understand how to optimise the parameters or incorporate effective techniques when necessary to maximise reconstruction accuracy. The subsequent section will delve into the different methods to achieve this.

\subsection{Strategies for accurate GF reconstruction}\label{Section: Strategies for accurate GF reconstruction}

\subsubsection{Optimisation of diffuse field parameters} \label{Section: Parameter optimisation}
Signals recorded at the two locations inevitably contain unwanted noise, such as vibrations from the surrounding environment and electronic interference stemming from equipment, which may obscure the diffuse fields. Generally, it is desirable for the diffuse fields to exhibit larger amplitudes than the noise to achieve better reconstruction quality. This noise contributes to the generation of reconstruction noise during the cross-correlation process, and a higher level of reconstruction noise ultimately leads to a decline in the overall quality of the reconstruction. 

\begin{figure}[!t]
  \centering
  \includegraphics[width=0.7\textwidth]{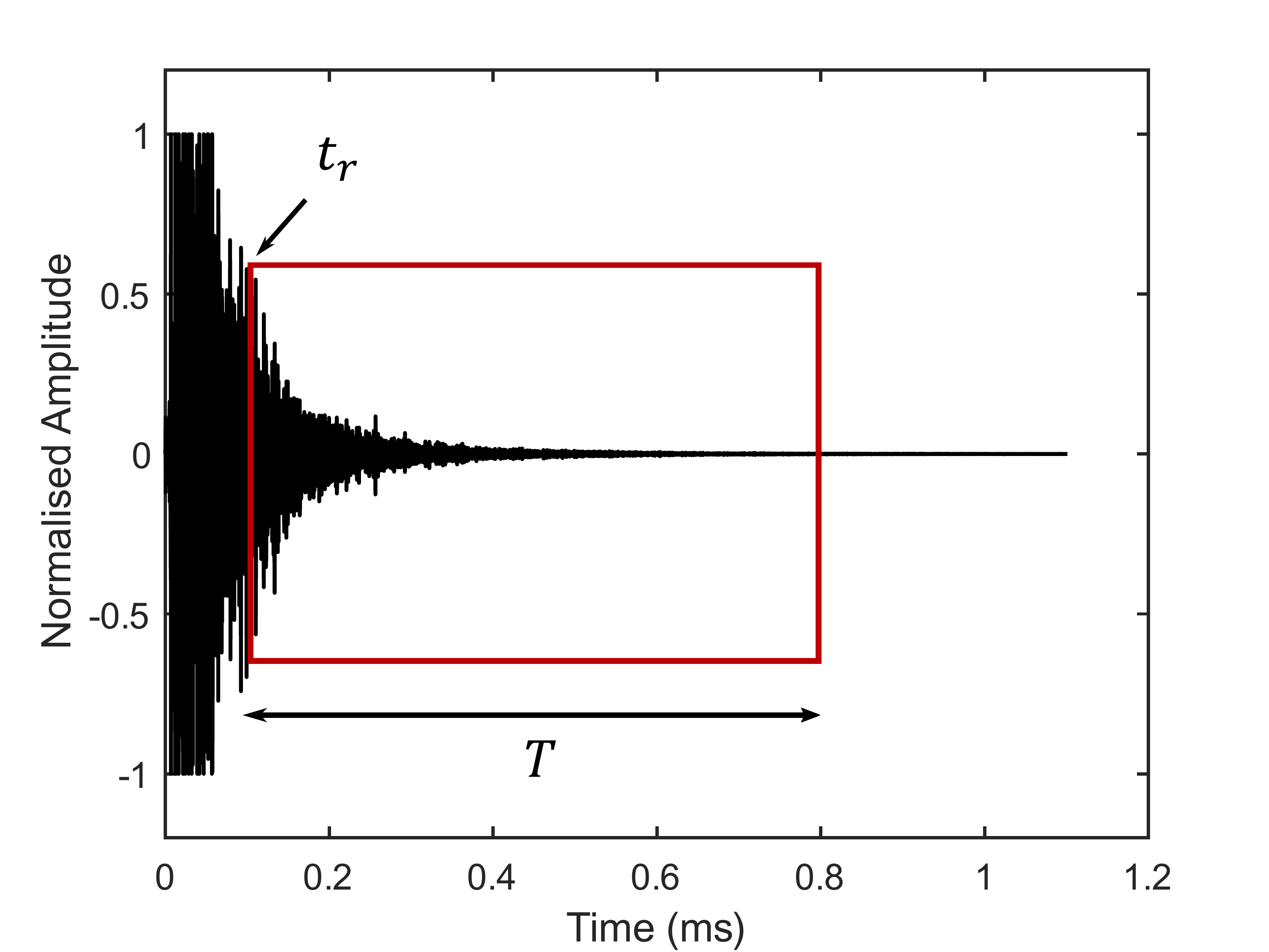}	
  \caption{A typical time-domain diffuse signal, characterised by two parameters: transmission delay $t_r$ and window length $T$, used to define the region of the diffuse field for GF reconstruction.}
  \label{fig:Diffuse signal}
\end{figure} 

Figure \ref{fig:Diffuse signal} shows how a typical time-domain diffuse signal looks like and the recording time needed. This longer time scale of milliseconds is usually used for diffuse field measurements in comparison to the microseconds used for typical NDE signals. The diffuse field used for GF reconstruction in this work is obtained after a transmission delay $t_r$ and within a window length $T$ as seen in Figure \ref{fig:Diffuse signal}. By optimising $t_r$, $T$ and source averages $N$ (i.e. the number of independent source locations), reconstruction noise can be reduced to achieve better reconstruction quality.

For the selection of $t_r$, it should be the time when the field becomes diffuse. This parameter can be approximated using the equipartition characteristic of a diffuse field \cite{Weaver1982OnMedia}, which involves ensuring uniform energy levels across different locations of the sample block. This assessment can be conducted through experiments or simulations by verifying that the Root Mean Square (RMS) of particle displacements at various locations remains consistent. Notably, the time when the diffuse field is formed is contingent upon the diffusivity of the material, which, in turn, is influenced by factors such as grain size and anisotropy.
 
For the selection of $N$ and $T$, Lobkis and Weaver suggested theoretical estimates in \cite{Lobkis2001OnField} for the amount of averaging over source locations and window length needed before convergence towards GF is achieved. If the frequency bandwidth of GF to be recovered is around 1 MHz, which is our representative frequency of interest for materials characterisation and NDE, the estimates suggests values in the range of tens of thousands for $N$ and hundreds of milliseconds for $T$, which are impractical to achieve, and this will further increase with higher frequency. Nevertheless, it is recommended that averaging over more source locations $N$ and longer window length $T$ can improve the reconstruction quality. 

However, a longer $T$ results in longer computation times and, due to energy dissipation, the amplitude of diffuse fields at later times might become too small, making them indistinguishable from unwanted noise. Utilising these signals would not contribute to the reconstruction of the GF. Therefore, there is a time threshold beyond which these signals should not be used. Here, we recommend the use of a spectral energy density curve derived from a diffuse signal to determine this threshold for the optimisation of $T$. This curve illustrates the change in ultrasonic spectral energy over time and provides a more reliable indication of ultrasonic diffuse fields compared to the time-domain signal.

\begin{figure}[!t] 
  \centering
  \includegraphics[width=0.7\textwidth]{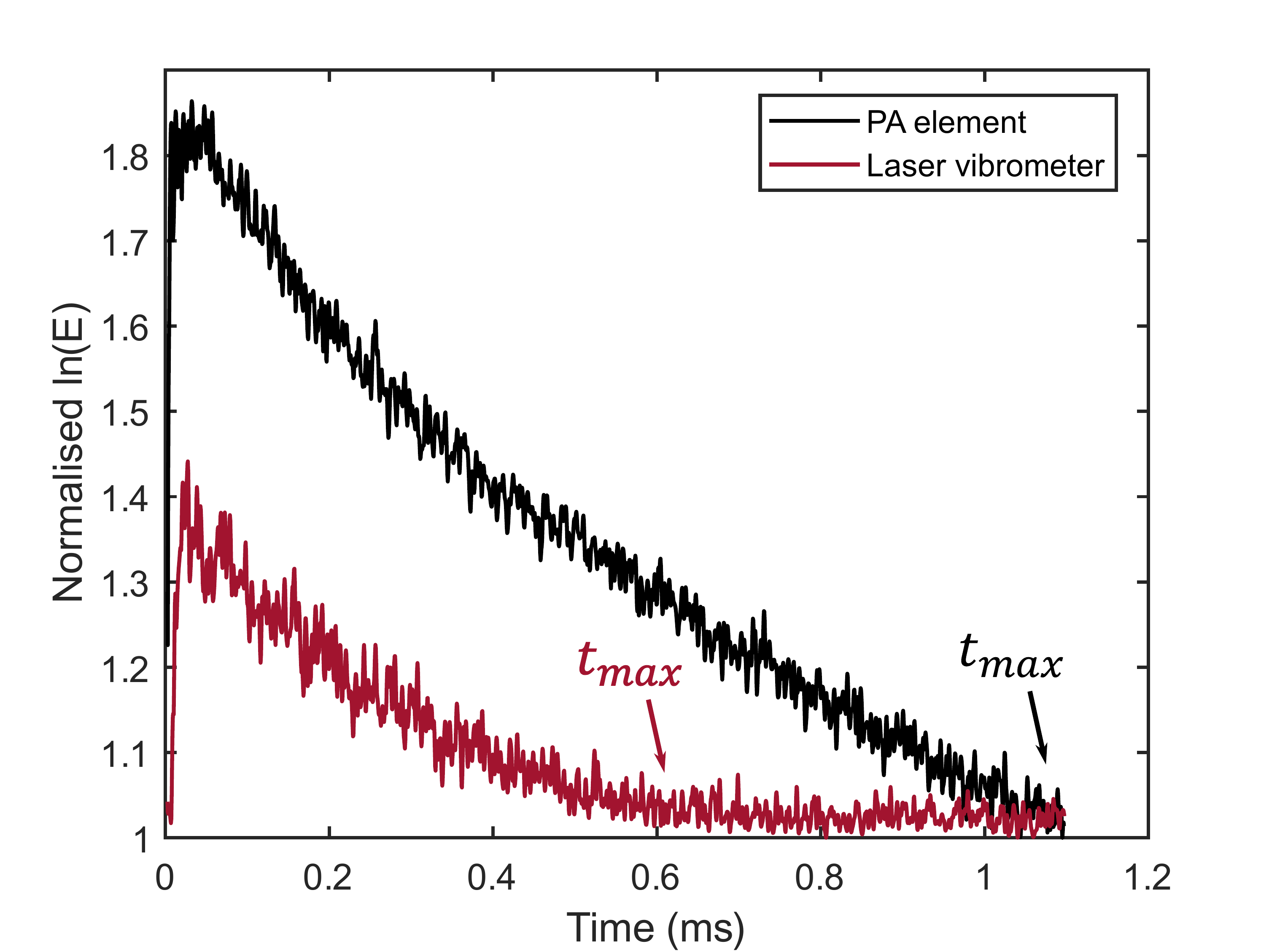}	
  \caption{Comparison of the energy density curves of signals received from a piezoelectric PA element (in black) and a laser vibrometer (in red).}
  \label{fig:Energy density curves}
\end{figure}

Conducting standard time-frequency analysis on a time-domain diffuse signal, as depicted in Figure \ref{fig:Diffuse signal}, allows for the generation of the corresponding spectral energy density curve shown in Figure \ref{fig:Energy density curves}. The process involves the following steps as seen in \cite{Weaver1998UltrasonicsFoam, Becker2003CharacterizationUltrasound, Lu2017CharacterizationTransducers}:
 
\begin{enumerate}
    \item Divide time-domain diffuse signal into overlapping time windows of length $\Delta t$ and apply a Hanning window on the signals within each window. In this study, $\Delta t$ is defined as 5 \textmu s, with a $90\%$ overlap between consecutive windows.
    
    The selection of $\Delta t$ is based on Equation 4 in \cite{Weaver1998UltrasonicsFoam}, which describes the inverse relationship between fluctuations of the measured energy density and the square root of the product of the overlapping time window length and frequency bandwidth. Specifically, $\Delta t = 5$ \textmu s corresponds to defining the frequency bandwidth as $\Delta f = 5.5$ MHz and allowing for energy density fluctuations of $20\%$. A larger window overlap of $90\%$ is chosen to ensure higher resolution in the analysis.

    \item Calculate the discrete-time Fourier transform (DTFT) for each time window and square it.

    \item Determine the spectral density of each time window in a certain frequency band by integrating the power spectral density in that bandwidth with $\Delta f$ centred around the central frequency $f_c$. In this study, $\Delta f$ is defined as $5.5$ MHz and $f_c$ as $5$ MHz.

    \item Calculate the logarithm of the spectral density of each time window and plot them against the time central of each respective time window.
\end{enumerate}

Figure \ref{fig:Energy density curves} shows the respective energy density curves for different receivers. The black curve corresponds to signals received from a piezoelectric PA element, while the red curve represents those received using a laser vibrometer. Both curves exhibit a similar shape, reflecting the ultrasonic energy density that experiences a sharp and rapid ascent to a peak before descending due to energy dissipation. Smaller amplitudes indicate smaller contributions from the corresponding diffuse fields to the reconstruction of the GF. The point at which the normalised amplitude reaches 1 signifies the time $t_{max}$ where unwanted noise begins to dominate the diffuse fields. Consequently, any signal used beyond this point will not contribute to the GF reconstruction. Overall, when selecting $T$, we recommend a larger value to improve reconstruction quality but the summation of $t_r$ and $T$ should never exceed $t_{max}$.
  
These curves also offer valuable insights into the difference in sensitivity of ultrasonic vibrations between a piezoelectric PA element and a laser vibrometer. A higher peak on the graph reflects better sensitivity to ultrasound signals, underscoring the comparatively limited sensitivity of the lasers, which would result in lower SNR of signals and consequently, poorer reconstruction accuracy when lasers are utilised. However, the decision regarding which type of receiver to use is not solely based on SNR. Despite the higher SNR associated with the contact transduction method, it requires the use of a couplant to excite ultrasound in the material. This presents a significant obstacle to accurate velocity measurement, as the thickness of the couplant is uncontrollable. Even small uncertainties on the micron scale can result in velocity measurement errors exceeding 150 ms\textsuperscript{-1}, which is unacceptable, especially in contexts such as crystallographic texture analysis, as demonstrated by Lan \textit{et al}. \cite{Lan2018DirectWaves}. On the contrary, laser interferometry offers the advantage of being a non-contact method, allowing for remote measurements, even under challenging conditions like elevated temperatures during manufacturing. This opens up significant possibilities for measurements across various manufacturing steps. Overcoming the inherent challenge of lower sensitivity is crucial for achieving this, and techniques to address this issue will be discussed in Section \ref{Section: Laser experimental procedure}.

\subsubsection{Deconvolution of a source-dependant factor} \label{Section: Deconvolution of source-dependant factor}

As mentioned in Section \ref{Section: GF Reconstruction theories}, the diffuse field correlation contains a source-dependent factor that requires deconvolution to achieve accurate GF reconstruction. In this study, we employed the Time Reversal (TR) symmetry interpretation of GF reconstruction, as used in \cite{Derode2003RecoveringL, Larose2004ImagingFields}, to determine this factor $F(-\omega)$:

\begin{equation}\label{eq:2}
  F(-\omega) = W(-\omega) \cdot S(-\omega) \cdot S(\omega) \cdot G_{S,S}(-\omega) = \hat{T}F(\omega)
\end{equation}
 
where $\omega$ represents the angular frequency, $W(\omega)$ represents the excitation waveform, $S(\omega)$ and $S(-\omega)$ are, respectively, the transfer function of the source used to generate the diffuse field and its time-reversed form, $G_{S,S}(\omega)$ denotes the GF at source location S, $\hat{T}$ is the TR operator, and $\hat{T}F(\omega)$ is an alternative representation of the source-dependent factor. 
 
Derivation for $F(-\omega)$ can be found in Appendix \ref{Appendix: Definition of source-dependent term for deconvolution} and is based on our aim of reconstructing the coherent response, which is the convolution of the GF with the excitation pulse. However, if the objective is to reconstruct the GF, the factor to be removed would be $F(-\omega) \cdot W(\omega)$ as mentioned by Derode \textit{et al.} \cite{Derode2003RecoveringL}. The derivation assumes a single source, so if multiple sources are utilised for averaging, $F(-\omega)$ will represent the averaged value. Assuming $S(-\omega)$ approximates the receiver transfer function, $F(\omega)$ which is represented by $W(\omega) \cdot S(\omega) \cdot S(-\omega) \cdot G_{S,S}(\omega)$, describes the pulse-echo response of the source. Therefore, the source-dependent factor $\hat{T}F(\omega)$ represents the time-reversed pulse echo response of the source. 

\begin{figure}[!t]
    \centering
    \includegraphics[width=1\textwidth]{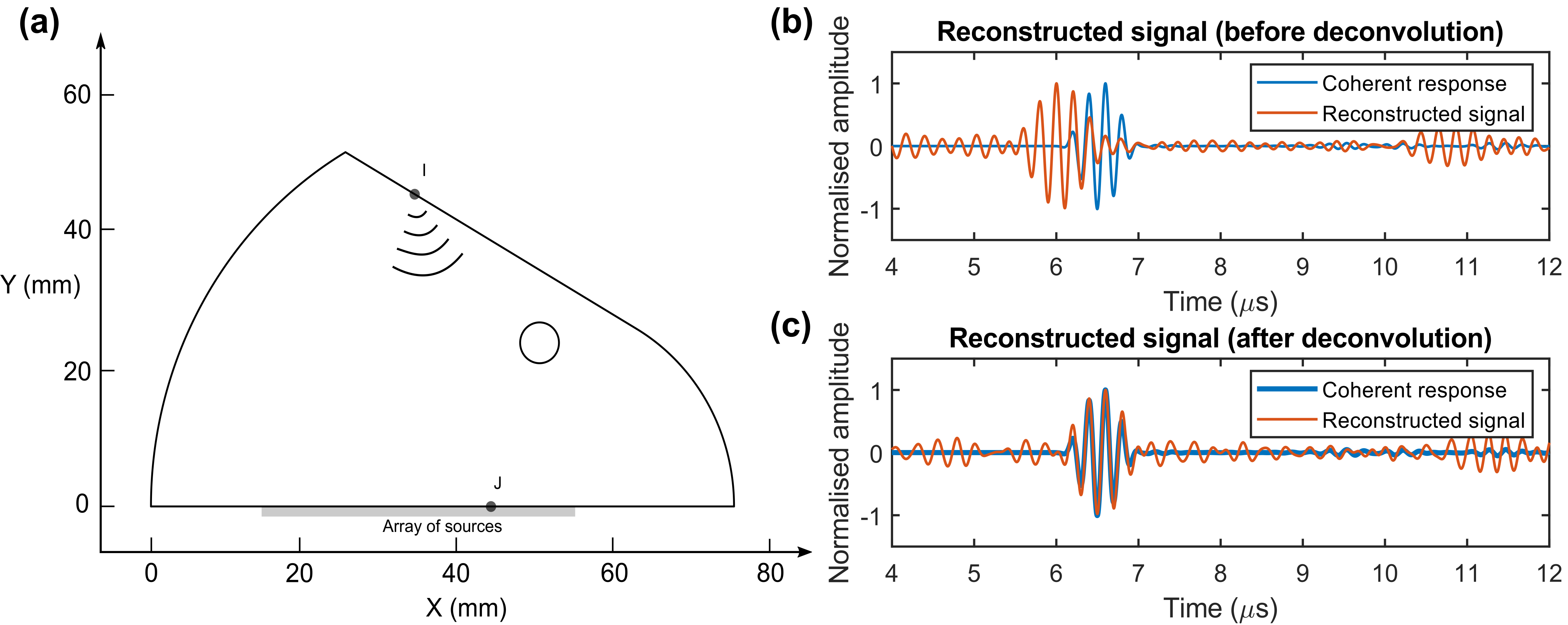}	
    \caption{FE model to simulate the GF reconstruction process from diffuse fields. \textbf{a} shows the schematic of the sample modelled using FE. Diffuse fields were collected at location I and J to reconstruct the coherent response received at J from I. \textbf{b} compares the reconstructed signal as a result of only cross-correlation against the coherent response. \textbf{c} compares the reconstructed signal as a result of both cross-correlation and deconvolution against the coherent response.}
    \label{fig:Deconvolution result (simulation)}
\end{figure}

In order to gain more understanding about the reconstruction process from diffuse fields, especially the deconvolution step, a Finite Element (FE) model was generated using the high-speed GPU-based software Pogo \cite{Huthwaite2014AcceleratedGPU}. The model replicated a ISO 7963 miniature calibration block \cite{TMTeckOnBlock}, and its schematic is illustrated in Figure \ref{fig:Deconvolution result (simulation)}a. Diffuse fields that were generated by an array of 64 sources, were collected at locations I and J and averaged to reconstruct the coherent response between the two locations. Each source was modelled through a force applied perpendicular to the surface, to nodes within a 0.5 mm aperture. The nodes were forced using a 5-cycle Hann-weighted sine wave, with a central frequency of 5 MHz.

The cross-correlation of diffuse fields received at locations I and J was calculated in the time domain and compared with the corresponding coherent response between the two locations. The comparison between both signals is illustrated in Figure \ref{fig:Deconvolution result (simulation)}b. The main feature of the direct pitch-catch waveform which is the first arrival is clearly present in the correlation as seen from the orange signal. However, it is also apparent that the reconstructed first arrival occurs at a different time compared to the coherent response depicted in blue and appears to have a different waveform shape. Notably, the first arrival pulse occurs roughly 1 \textmu s earlier in time in this case. This equates to a velocity difference of about 700 ms\textsuperscript{-1} which will render velocity measurement based on arrival time to be extremely inaccurate. This discrepancy arises not from a delay in the coherent response but due to the presence of the source-dependent factor $\hat{T}F(\omega)$.

Therefore, the deconvolution of $\hat{T}F(\omega)$ is necessary for accurately reconstructing the coherent response. Referring to Equation \ref{eq:2}, $F(\omega)$ can be simplified to $W(\omega) \cdot G_{ss}(\omega)$ in this case, given the absence of transducer characteristics, as excitation was defined by applying forces on the source nodes and responses were retrieved by recording the displacements of the receiver nodes. Since $W(\omega) \cdot G_{ss}(\omega)$ describes the pulse-echo response, it was obtained as the signal retrieved after designating the nodes for excitation as receivers. Following this, a TR operation was applied before converting the signal to the frequency domain to obtain $\hat{T}F(\omega)$. Deconvolution was then performed by transforming the diffuse field correlation to the frequency domain to obtain $R_{j,i}(\omega)$ and dividing it by $\hat{T}F(\omega)$, as seen in Equation \ref{eq:1}. The outcome, represented in orange, is juxtaposed with the coherent response in Figure \ref{fig:Deconvolution result (simulation)}c. The significantly improved agreement with the coherent response clearly highlights the importance of the deconvolution step in attaining an accurate estimation of the coherent response.
 
It is crucial to emphasise that the complexity of obtaining $F(\omega)$ becomes more complex in experimental GF reconstruction, as it becomes dependent on the properties of the source transducer and the setup. Consequently, establishing a reliable method to account for and accurately determine $F(\omega)$ becomes imperative. This will be addressed in Section \ref{Section: Experimental implementation of GF reconstruction}, which covers the experimental implementation of this method.

\subsubsection{One-bit Normalisation} \label{Section: One-bit Normalisation}
As discussed in Section \ref{Section: Parameter optimisation}, the amplitude of diffuse fields decreases over time due to energy decay, resulting in lower-amplitude signals that contribute less to the GF reconstruction process \cite{Weaver2003ElasticPhonons}. This phenomenon reduces the effective window length for reconstruction, consequently impacting the accuracy of the reconstruction process. To address this issue, temporal normalisation techniques can be employed to mitigate the impact of diminishing signal amplitudes over time.

Temporal normalisation is a commonly applied procedure in the preprocessing of seismic noise data before conducting cross-correlation in seismology \cite{Campillo2003Long-RangeCoda, Shapiro2004EmergenceNoise, Sabra2005ExtractingNoise}. Simply performing a basic cross-correlation between noise records from two stations can disproportionately emphasise the most energetic segments of the noise, such as signals from unanticipated earthquakes, or down-weigh the less energetic ones. The use of temporal normalisation aims to reduce the influence from such non-uniform noise sources on seismic diffuse fields (e.g. ambient noise, scattered coda waves) utilised for reconstructing GF. Here, we see the applicability of this method to mitigate the energy decay issue since its fundamental mechanism is to give more weights to the smaller-amplitude signals. 

\begin{figure}[!t]
  \centering
  \includegraphics[width=1\textwidth]{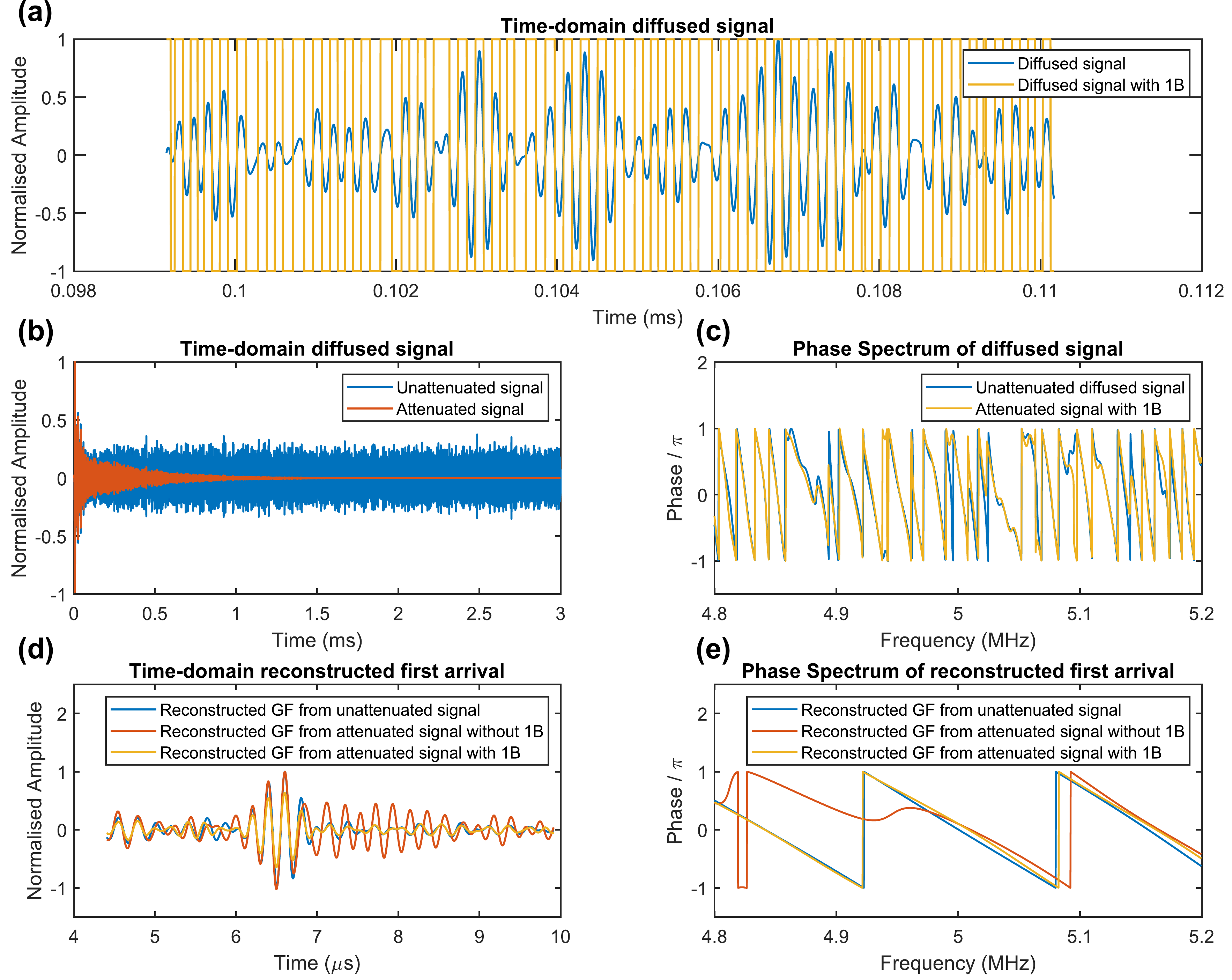}	
  \caption{Simulated diffuse signals, along with reconstructed GF, and their phase information with 1B normalisation. \textbf{a} overlays the signals before and after 1B normalisation. \textbf{b} displays two initially identical signals, with one experiencing attenuation. \textbf{c} compares the phase spectrum of the unattenuated signal and the binary form of the attenuated signal in \textbf{b}. \textbf{d} compares the reconstructed first arrival from the unattenuated signal, attenuated signal and the binary form of the attenuated signal in \textbf{b}. \textbf{e} compares the phase spectrum of the reconstructed first arrivals seen in \textbf{d}.}
  \label{fig:BQ explanation}
\end{figure}

The study by Bensen \textit{et al.} explores five main temporal normalisation methods, including one-bit (1B) normalisation and running-absolute-mean normalisation \cite{Bensen2007ProcessingMeasurements}. The latter's implementation is detailed in the paper, with 1B actually being a simplified form of this method that retains only the signs of the signals, as seen in Figure \ref{fig:BQ explanation}a. Despite its simplicity, several studies have demonstrated that 1B effectively eliminates spikes in seismic noise and enhances the SNR of cross-correlation functions \cite{Campillo2003Long-RangeCoda, Shapiro2005High-ResolutionNoise, Larose2007ReconstructionForcing}. Therefore, we opted for 1B  due to its straightforward implementation and documented efficacy in enhancing SNR, as also demonstrated in acoustic experiments conducted by Larose \textit{et al.} \cite{Larose2004ImagingFields}. 
 
To fundamentally grasp the workings of 1B normalisation and its impact on GF reconstruction in our experimental setup, we offer a simulation-based demonstration using the signals seen in Figure \ref{fig:BQ explanation}. All signals within Figure \ref{fig:BQ explanation} were generated using the same FE model that was introduced in Section \ref{Section: Deconvolution of source-dependant factor}. In Figure \ref{fig:BQ explanation}b, two signals are presented: the blue signal represents an unattenuated diffuse signal, whereas the orange signal undergoes dissipation. This dissipation is introduced by incorporating a mass proportional damping coefficient of 6000 s\textsuperscript{-1} when specifying the material properties of the model. This step is undertaken to emulate the energy decay encountered in real-life experiments. Moving to Figure \ref{fig:BQ explanation}c, the phase spectrum of the unattenuated diffuse signal and that of the attenuated signal, post-conversion, are displayed. The phase spectrum of the binary signals is shown to closely resemble that of the unattenuated signal, demonstrating that the phase information is effectively preserved through the conversion process. This preservation of phase information extends to the resultant GF emerging from the correlation, as evident in both Figure \ref{fig:BQ explanation}d and Figure \ref{fig:BQ explanation}e where the time-domain first arrivals and their phase spectrum respectively exhibit good agreement with each other. This observation aligns with the understanding that one-bit noise correlation has the same phase as the raw noise correlation \cite{Cupillard2011TheNoise}. Both figures also serve to highlight the poorer reconstruction quality associated with shorter effective window length when 1B is not applied. As shown in Figure \ref{fig:BQ explanation}d, the first arrival waveform remains identifiable but the SNR is lower due to the presence of larger amplitude reconstruction noise (in the region of 7 to 9 \textmu s) after the first arrival. Referring to its phase spectrum in Figure \ref{fig:BQ explanation}e, depicted in orange, it is evident that the phase spectrum does not agree well with the case when 1B is applied. 

The findings illustrate that 1B normalisation enables decayed signals with lower amplitudes to make a more substantial contribution to GF reconstruction, thereby accelerating the convergence toward GF reconstruction within the confines of the same parameter $T$. This highlights its significance in our experimental implementation of GF reconstruction especially for low-SNR scenarios such as when laser receivers are used.

It is important to acknowledge that 1B normalisation introduces quantisation errors at higher frequencies, mainly due to the sharp edges of the square pulses. These errors manifest as slightly increased amplitudes of high-frequency components, extending beyond the frequency bandwidth of interest, when 1B is applied compared to its absence. In our simulations, these errors were noticeable from frequencies as low as around 10 MHz up to 100 MHz while our bandwidth of interest is 3.25 to 7.25 MHz. However, it is worth noting that these amplitude variations are relatively small and generally do not affect the accuracy of the reconstructed GF or velocity measurement results. To further ensure that the reconstructed GF remains focused on the desired frequency range relevant to the velocity measurements, low- or band-pass filters can be incorporated based on the specific bandwidth of interest.

\subsection{Group or phase velocity?} \label{Section: Group or phase velocity?}
To conduct velocity measurement from the reconstructed GF, it is first important to understand what type of velocity we are measuring, i.e. group or phase. These two are the same for isotropic non-dissipative materials but can be drastically different for anisotropic or strongly absorbing materials. Our interest here is in anisotropic materials, normally metals with low absorption. Therefore, we choose to consider the GF for an anisotropic medium where the group and phase velocity differs in different directions. 

In this study, we will compare a theoretical model of 3-D GF in anisotropic material with the theoretical calculation of both the group and phase velocity from Christoffel's equation under the plane wave assumption. For the theoretical model, we refer to the 3-D GF for anisotropic solids presented by Wang and Achenbach \cite{Ang1995Three-dimensionalSolids}. In the paper, the 3-D GF corresponds to the displacement field in the $x_p$ direction produced by an impulsive point load applied at the origin in the $x_k$ direction. The expression for the 3-D frequency domain GF shown in \cite{Ang1995Three-dimensionalSolids} is:

\begin{equation}\label{eq:4}
\begin{split}
  g_{pk}(\boldsymbol{x}, \omega) = \frac{i}{8\pi^2}\int_{|\boldsymbol{n}|=1}\sum_{m=1}^{3}\frac{k_m E_{pm}E_{km}}{2\rho c_m^2}e^{ik_m|\boldsymbol{n} \cdot \boldsymbol{x}|} dS(\boldsymbol{n}) \\
  + \ \frac{1}{2\pi^2} Re\int^{\pi}_{0}\sum_{m=1}^{3}[\frac{A_{pk}(\boldsymbol{k})}{\partial_{k_3}D(\boldsymbol{k})} \frac{sgn(x_3)}{\boldsymbol{k} \cdot \boldsymbol{x}}]_{k_3 = k^m_3} d\theta
\end{split}
\end{equation}

\noindent where $\boldsymbol{x}$ is the location of the receiver, $\rho$ is the density of the material, $m = 1, 2, 3$ represents the three different wave modes. For each wave mode, $c_m$ is the phase velocity and $k_m$ is the wave number, $E_{pm}$ and $E_{km}$ corresponds the polarisation amplitude in the $x_p$ direction and $x_k$ direction respectively. Other terms in the equation, such as $A_{pk}(\boldsymbol{k})$ and $D(\boldsymbol{k})$, requires the use of Christoffel tensor. Further details on solving for these terms can be found in the paper.

The medium selected was a single crystal cubic steel with density of 7850 kgm\textsuperscript{-3}. The elastic constants used to define the medium are given in units of GPa, by \cite{Talling2008DeterminationMetal}:

\begin{gather*}\
  c_{11} = 21.83, \, c_{12} = 21.75, \, c_{44} = 11 
\end{gather*}

\begin{figure}[!t]
  \centering
  \includegraphics[width=1\textwidth]{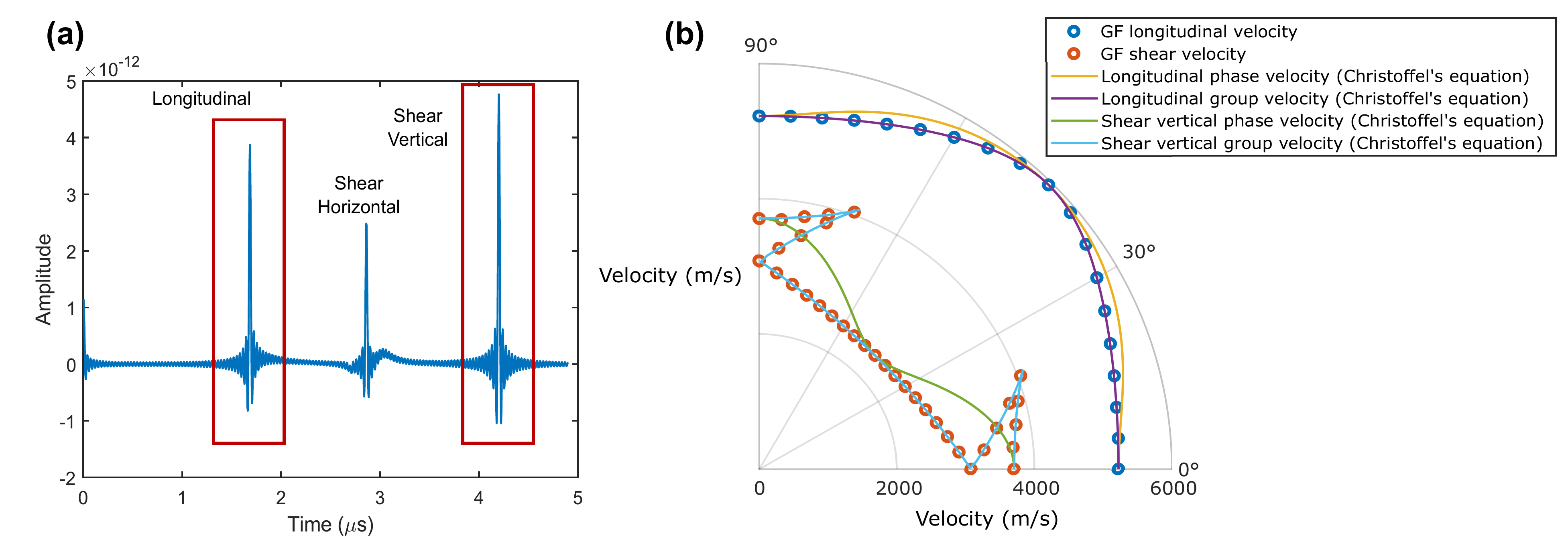}
  \caption{Velocity measurement of longitudinal and shear vertical wave modes from GF. \textbf{a} shows an example of the time-domain GF after undergoing inverse Fourier Transform from the corresponding 3-D GF in the frequency domain. \textbf{b} compares the velocities of different wave modes calculated using arrival times extracted from time-domain signals, similar to the one in \textbf{a}, in different directions, to the respective group and phase velocity values from Christoffel's equation.}
  \label{fig:Velocity measurement of GF}
 \end{figure}

In this medium, notable variations exist in the group and phase velocities of longitudinal and shear wave modes across different directions, with the differences between them spanning the range from zero to hundreds of ms\textsuperscript{-1}. This variability allows us to identify the type of velocity being measured. A circular array of receivers was defined in the XZ plane, positioned 10 mm away from the origin source, with each receiver separated by $5^\circ$. To optimise computation time, only the first quadrant (i.e., $0$ to $90^\circ$) was considered, leveraging the four-fold symmetry axes inherent in a cubic crystal structure. By employing Equation \ref{eq:4}, we calculated the GF in the frequency domain across these 19 directions. Subsequently, the results were transformed into time-domain signals using inverse Fourier Transform. An illustrative result is depicted in Figure \ref{fig:Velocity measurement of GF}a. From such signal, we can distinguish the different wave modes and extract their respective arrival times to calculate the velocities. It is noteworthy from Figure \ref{fig:Velocity measurement of GF}a that the shear horizontal mode was present; however, as the group and phase velocities of this wave mode remain constant across all directions in this plane, it was not considered in the analysis.

In Figure \ref{fig:Velocity measurement of GF}b, the velocity values are superimposed with the longitudinal and shear vertical group and phase velocities obtained through Christoffel's equation. It is important to highlight that all phase velocity values are plotted according to their respective phase velocity directions, while group velocity values are plotted in accordance with their group velocity directions. The close correspondence between the values computed using the 3D GF equation and the group velocity for both longitudinal and shear modes establishes that the velocity measured from GF corresponds to the group velocity. This observation highlights that when utilising a point-to-point measurement method where one point serves as a source and another as the reception to obtain the GF, the measured velocity is the group velocity.

While the samples employed in this paper are lossless and isotropic, exhibiting identical group and phase velocities in all directions, we will specifically designate the measured velocity as the group velocity for clarity.

\section{Experimental implementation of GF reconstruction} \label{Section: Experimental implementation of GF reconstruction}
This section describes the implementation of GF reconstruction for velocity measurements using different types of receivers, namely PA and laser receivers. The experiment that uses PA as both a source and receiver is prioritised as the initial step due to its ability to generate coherent signals for validation and its higher receiver sensitivity as compared to lasers. This approach allows us to establish the experimental configuration, data acquisition, and signal processing needed before applying the methodology to the more challenging implementation of using laser receivers for velocity measurement on a sample with curved surfaces.

\subsection{Contact PA receivers} \label{Section: PA experimental procedure}
To assess the feasibility of accurately reconstructing coherent responses at two arbitrary locations from diffuse fields and subsequently determining corresponding velocity values, a laboratory experiment was conducted. It utilised a calibration block, such as those used routinely by NDE practitioners, that was intentionally heat-treated to be homogeneous and isotropic. PA transducers were employed for their dual capability as both a source and a receiver. This allowed velocity measurements in various directions using both coherent signal and reconstructed GF where the coherent velocities were used for validation. Despite the need for couplant, which introduced measurement errors, the same transducer placement for both coherent and diffuse field measurements was used. This ensured that both measurements were affected by the couplant-related errors equally, so any differences in velocity between the methods were not due to couplant; thereby allowing for accurate validation of the reconstructed results. Furthermore, using transducers consisting of multiple elements enabled averaging over more source locations, which was crucial for achieving higher accuracy during the reconstruction process as discussed in Section \ref{Section: Parameter optimisation}. 

\begin{figure}[!t]
  \centering
  \includegraphics[width=0.47\textwidth]{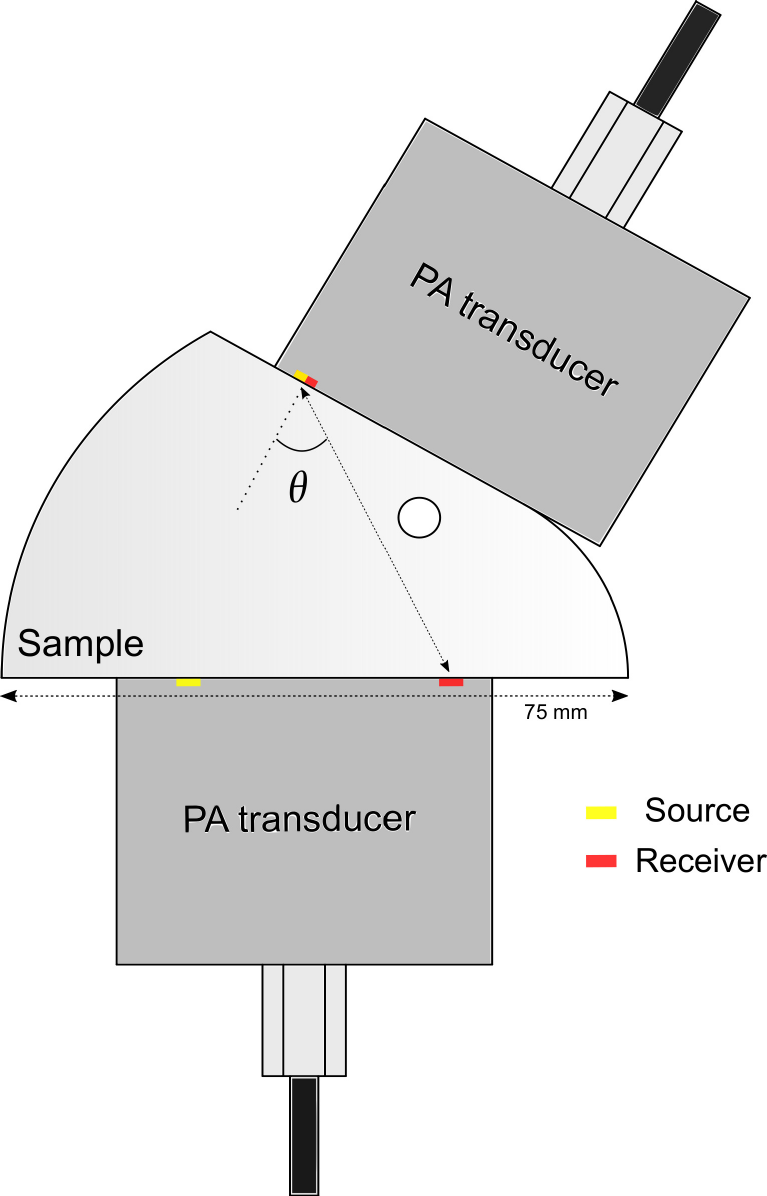}	
  \caption{Schematic of experimental setup for reconstructing and validating velocities measured in different directions with PA receivers}
  \label{fig:PA setup}
 \end{figure}
 
The schematic depicting the experimental setup can be seen in Figure \ref{fig:PA setup}. The sample used was a TB1065-1 ISO 7963 miniature calibration block from Olympus. It was made of 1018 Steel and had a thickness of 25 mm with a 0.5 mm diameter hole. The experiment was conducted using two Imasonic 64-element PA with nominal frequency of 5 MHz and pitch of 0.6 mm, interfaced with a Verasonics Vantage 128 array controller. The controller allowed a maximum excitation voltage of 96 V, which was utilised to enhance the SNR of the diffuse signals. Data points were captured at a sampling frequency of 25 MHz but were upsampled to 1 GHz for better resolution and therefore more accurate diffuse field correlation. Gel couplant was applied between the PA transducer and the sample surface.
 
In this specific experiment, we utilised 32 unique source-receiver and receiver-receiver pairs for the coherent and diffuse fields measurements respectively. Each of these 32 measurements corresponds to a specific measurement angle denoted as $\theta$. This angle was defined as the angle between the normal line extending from the PA element at the top and the direct line connecting the two PA elements, as illustrated in Figure \ref{fig:PA setup}. Using the same transducer placements, different PA configurations were used for coherent and diffuse field measurements. For coherent measurement, the configuration consisted of a source element at the top transducer along with 32 alternating receiver elements at the bottom transducer, indexed as 1st, 3rd, 5th, and so on, up to the 63rd element. For diffuse field measurements, the element at the top transducer was reconfigured to function as a receiver, and all 64 elements ($N = 64$) of the bottom transducer were employed as sources to generate diffuse fields. For the cross-correlation calculation, the diffuse fields were defined with a value of ${t_r}$ set at 0.1 ms and a $T$ of 300 \textmu s, following the procedures outlined in Section \ref{Section: Parameter optimisation}.

\begin{figure}[!t]
  \centering
  \includegraphics[width=0.7\textwidth]{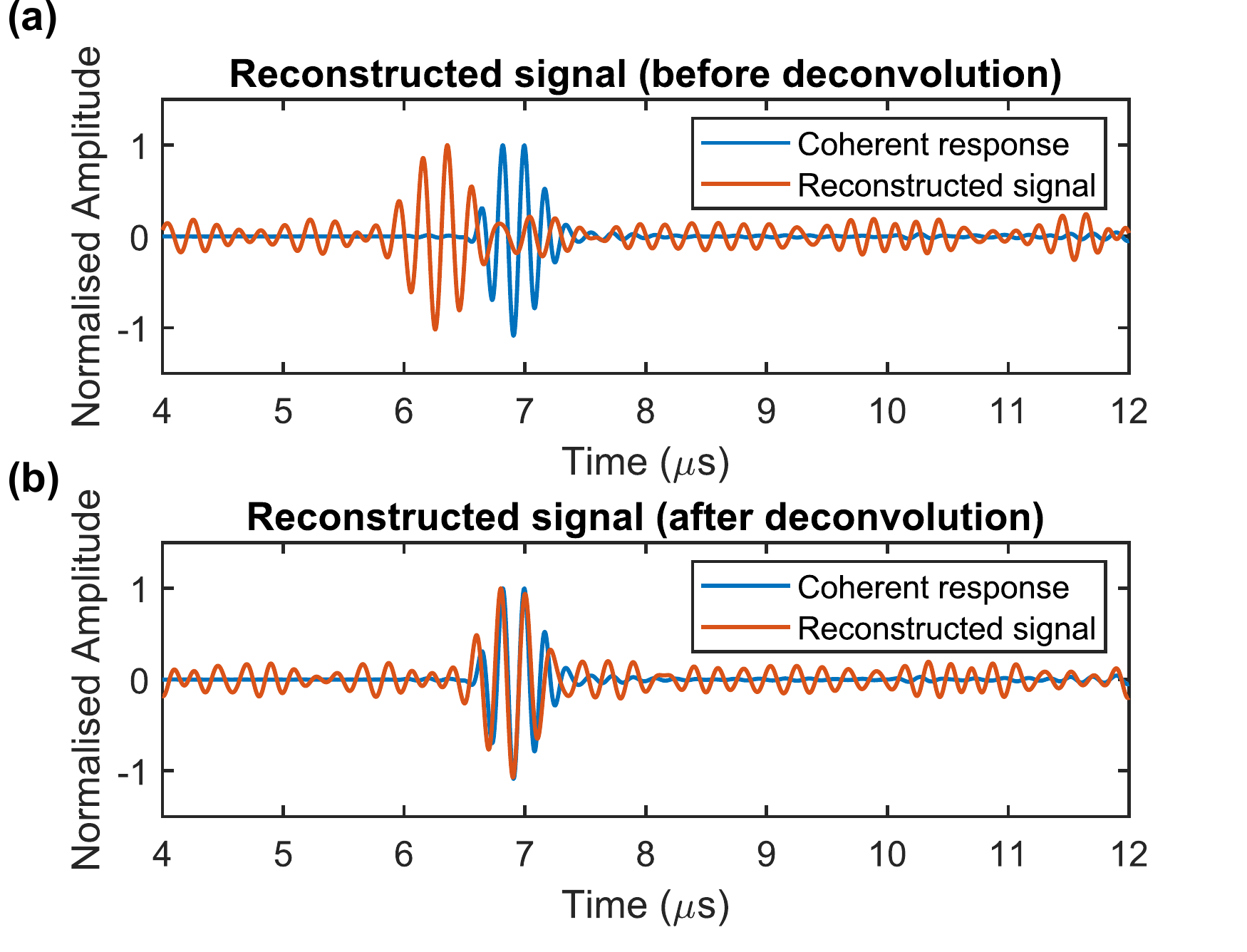}	
  \caption{Coherent response of the 20th receiving PA element and the respective reconstructed signal. \textbf{a} compares the reconstructed signal as a result of only cross-correlation against the coherent response. \textbf{b} compares the reconstructed signal as a result of both cross-correlation and deconvolution against the coherent response.}
  \label{fig:Deconvolution result}
\end{figure}
 
The time-domain diffuse field correlation was first calculated at all 32 measurement angles. Figure \ref{fig:Deconvolution result}a shows the comparison of the reconstruction signals against the corresponding coherent response at one measurement angle. The result is similar to that in Figure \ref{fig:Deconvolution result (simulation)}b because $\hat{T}F(\omega)$ has yet to be deconvolved.

\begin{figure}[!t]
  \centering
  \includegraphics[width=0.8\textwidth]{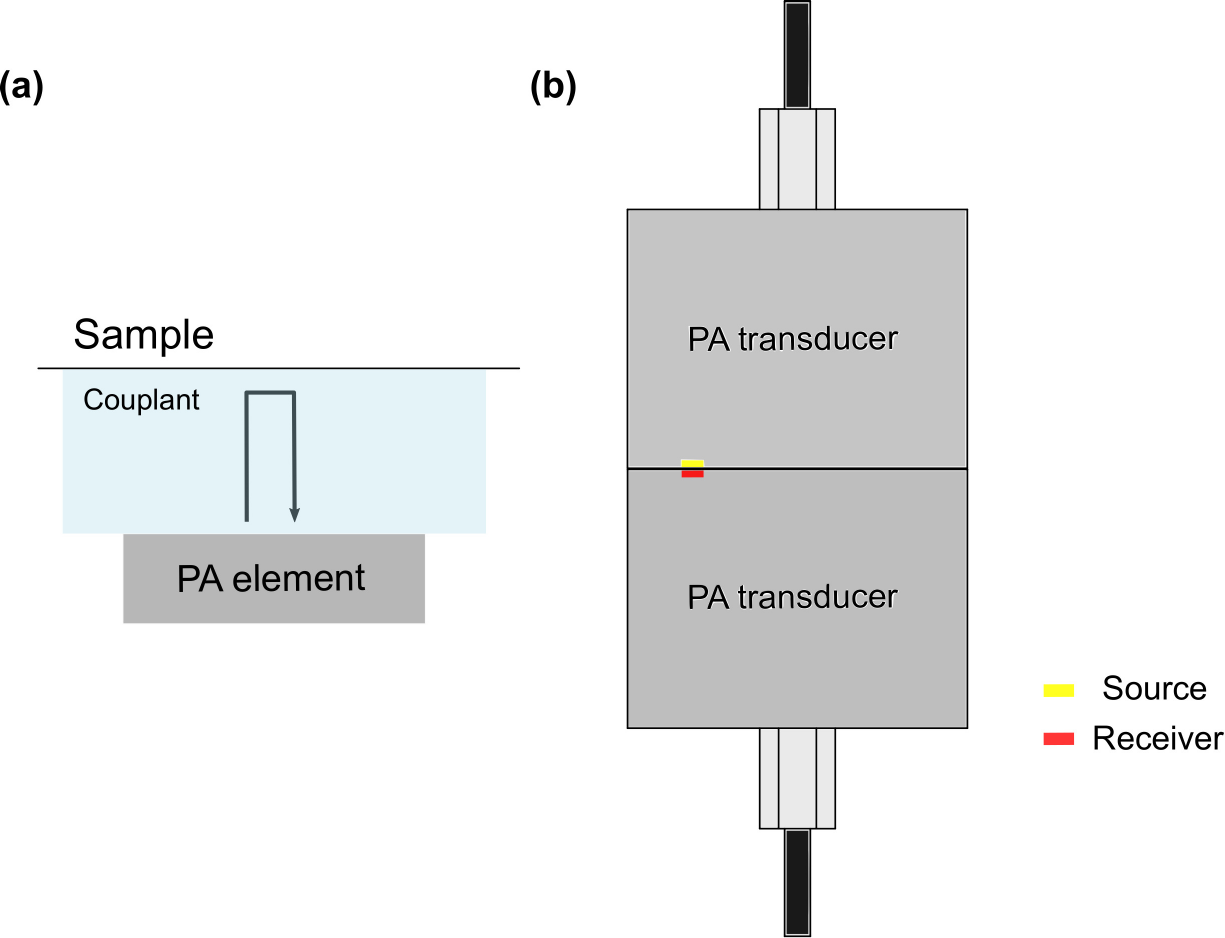}	
  \caption{Experimental considerations for defining the source-dependent factor that requires deconvolution for accurate Green's Function (GF) reconstruction. \textbf{a} provides a close-up view of the propagation path by the excitation waveform by a PA element to define the source-dependent factor in the experimental setup. \textbf{b} shows the setup required to approximate the source-dependent factor.}      	\label{fig:Deconvolution experimental setup}
\end{figure}

Expanding on the deconvolution step introduced in Section \ref{Section: Deconvolution of source-dependant factor}, it was essential to recognise the distinctions in its implementation between simulations and experiments, specifically in acquiring $F(\omega)$. Unlike the FE model where transducer characteristics were not present, experiments required the consideration of the transducer functions $S(\omega)$ and $S(-\omega)$. Ideally, all terms in $F(\omega)$ can be easily obtained from the early-time pulse echo response, which includes the reflected pulse from the sample boundary as shown in Figure \ref{fig:Deconvolution experimental setup}a. However, this reflected pulse is often obscured by the element ring down. Instead, we approximated the pulse-echo response simply as a phase-reversed excitation signal, with the phase reversal resulting from reflection at the sample boundary. To obtain the excitation signal, while taking into account the transducer characteristics $S(\omega)$ and $S(-\omega)$, we used the configuration depicted in Figure \ref{fig:Deconvolution experimental setup}b. The setup involved positioning two identical transducers, which were used as sources in the diffuse field measurement, facing each other, with one functioning as a transmitter and the other as a receiver. The excitation pulse used had to be identical to the one employed during the measurement. The resultant measured signal was the excitation signal and can be described by $W(\omega) \cdot S(\omega) \cdot S(-\omega)$, by assuming that $S(-\omega)$ represented the receiver function. Therefore, the only adjustment required to obtain $W(\omega) \cdot S(\omega) \cdot S(-\omega) \cdot G_{S,S}(\omega)$ is the application of a phase reversal on the signal. Subsequently, a time reversal operation was applied to obtain $\hat{T}F(\omega)$. This procedure ensured an accurate representation of the experimental conditions, which is important since $\hat{T}F(\omega)$ depended on these conditions.

However, the inevitable presence of reconstruction noise posed a challenge in applying deconvolution, given its high sensitivity to noise. To tackle this, we implemented the Wiener filter approach for deconvolution \cite{Levinson1946ThePrediction, Cicero2009PotentialMonitoring} in this study. This filter was designed to minimise the mean squared error between the estimated signal and the true signal, ensuring the best balance between bandwidth and SNR. In the frequency domain, the Wiener deconvolution can be expressed as \cite{Cicero2009PotentialMonitoring}:
 
\begin{equation}\label{eq:5}
  \Tilde{X}(\omega)=\ \frac{Z^{*}(\omega) Y(\omega)}{|Z(\omega)|^2 + \lambda}
\end{equation}

\noindent where $\Tilde{X}(\omega)$ is the reconstructed coherent response,  $Z(\omega)$ is the source-dependent factor or $\hat{T}F(\omega)$, $Y(\omega)$ is the diffuse field correlation and $\lambda$ is a regularisation factor used to avoid ill condition division in the determination of the filter and to mitigate the presence of high-frequency noise artefacts that might arise after the deconvolution.
 
A suitable regularisation factor was empirically determined to improve, or at very least not degrade, the signal quality by initially using the commonly employed value of $10^{-2}|Z(\omega)|^2_{max}$ \cite{Neal1993FlawInformation} as a reference value before proceeding with an iterative process. The result after deconvolution is shown in Figure \ref{fig:Deconvolution result}b. It was evident that the reconstructed first arrival now agreed more closely with the coherent one. There were still slight discrepancies in the waveform but these could be due to insufficient averages of different source locations or inaccurate estimation of transducer functions in the definition of $\hat{T}F(\omega)$ or both.

Following this, the velocity was computed at all 32 measurement angles using both coherent and reconstructed first arrivals, and a comparison was conducted. Velocity could be simply calculated by dividing the product of the distance between the source and receiver and the angular frequency by the phase difference between the transmitted and received pulses. In our calculations, the phase difference was extracted from the Fourier Transform of both signals. The distances between each source-receiver pair were readily known because the PA transducers were aligned with reference marks on the sample, each of which had a predetermined position \cite{TMTeckOnBlock}. This alignment allowed for the accurate determination of the distances used in the velocity calculations.

\begin{figure}[!t]
  \centering
  \includegraphics[width=0.7\textwidth]{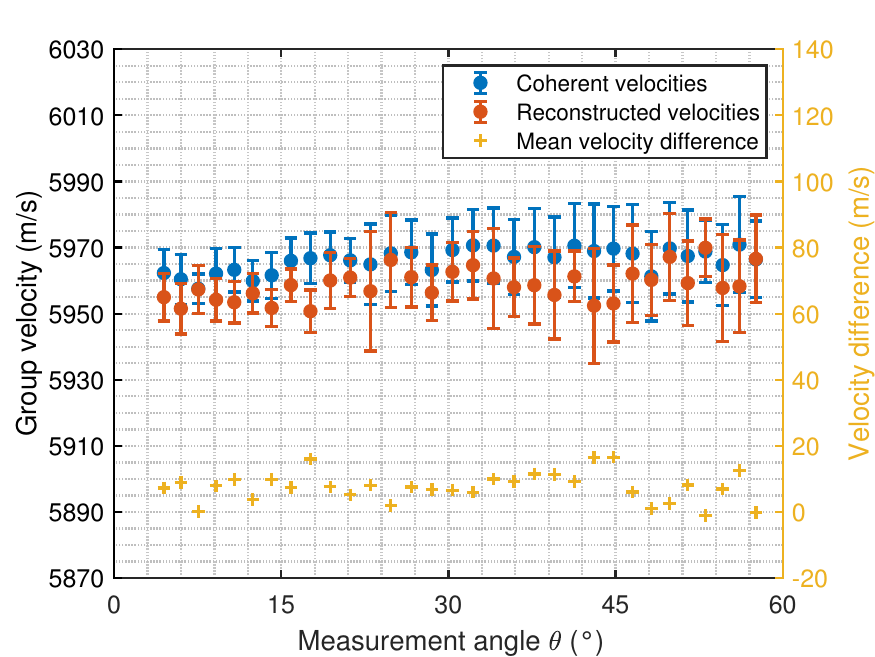}	
  \caption{Comparison of group velocity measured from coherent signals (in blue) and reconstructed signals (in orange) at 32 different measurement angles. The yellow data points (with axis scale on the right hand side) represent the difference in mean velocity between the coherent and GF reconstruction method at the corresponding measurement angle.}
  \label{fig:Measured phase velocities}
\end{figure}  

Twenty repetitions of each set of 32 velocity measurements were conducted, and the mean values, along with standard deviations represented by error bars, were calculated for every measurement angle. The outcomes are visually depicted in Figure \ref{fig:Measured phase velocities}. Additionally, the figure illustrates the calculated difference in mean velocity between both methods at the corresponding angles. The mean velocity difference fluctuates within the range of -2 to 17 ms\textsuperscript{-1}, with a mean value of 8 ms\textsuperscript{-1}. The standard deviations of individual velocities from both methods hover around 10 ms\textsuperscript{-1}, possibly due to fluctuations in experimental implementations. These findings illustrate a substantial resemblance in measured velocities when employing the GF reconstruction method, as opposed to relying on the coherent response. Notably, this similarity spans a wide angular range of normal incidence to 53$^{\circ}$, suggesting that the GF reconstruction method utilising diffuse fields provides an accurate approximation of coherent measurements at arbitrary locations. It is important to note that this applicability should extend to various types of receivers, not limited to PA transducers, and we will demonstrate a laser interferometer-based implementation in the next section.

\subsection{Non-contact laser receivers}\label{Section: Laser experimental procedure}

In this aspect of the study, the methodology is applied to laser receivers to measure velocity in different directions and the results are validated against the ground truth velocity of an isotropic block. The block was designed to have curved surfaces for velocity measurements and flat surfaces to allow the use of PA transducers as the source. This enables the generation of a diffuse field and allows averaging over more source locations to improve reconstruction accuracy.

\begin{figure}[!t]
  \centering
  \includegraphics[width=0.8\textwidth]{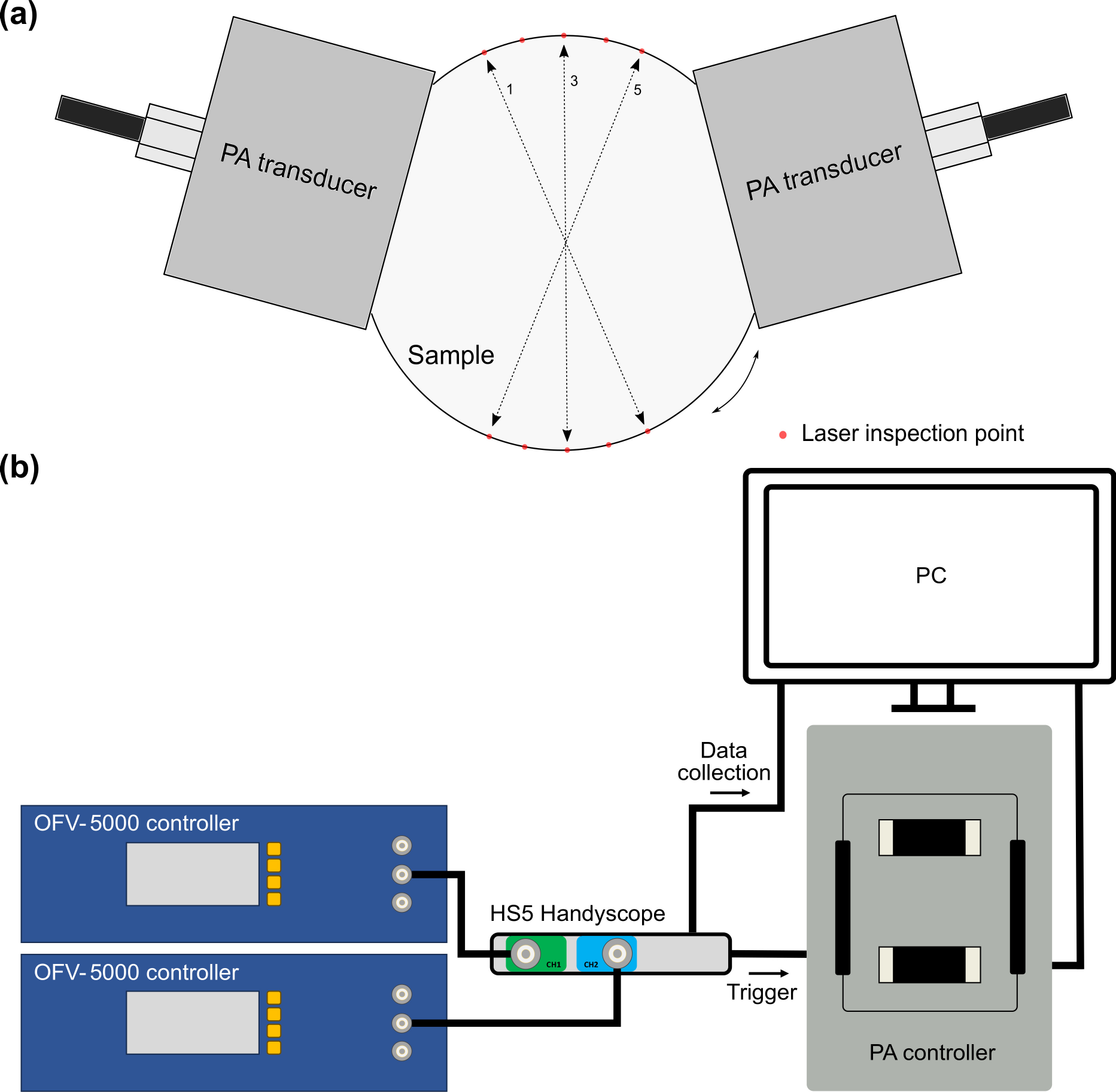}	
  \caption{Schematic of experimental setup for GF reconstruction with laser receivers. \textbf{a} illustrates the measurement setup by showing how measurements are conducted in five different directions using laser receivers. \textbf{b} depicts the synchronisation of the PA controller with the OFV-5000 vibrometer controller to ensure simultaneous excitation from the PA transducer and reception from the laser receivers.}
  \label{fig:Laser experimental setup}
\end{figure}

The experimental setup is illustrated in Figure \ref{fig:Laser experimental setup}. The sample used was an isotropic Aluminium 6082 block with a diameter of 70 mm and a thickness of 25 mm. In this set of experiments, the same PA transducers were used for excitation, while two Polytec OFV-505 laser vibrometer sensor heads were employed for reception. As depicted in Figure \ref{fig:Laser experimental setup}a, the sensor heads were positioned at opposite ends of the sample such that the distance between the two inspection points on the sample surface corresponds to the diameter of the sample. They each interfaced with an OFV-5000 vibrometer controller, facilitating the transfer of displacement data (determined using a DD-300 displacement decoder) from the controller to a computer via a HS5 Handyscope USB oscilloscope. The Handyscope also generated and sent a trigger signal to the PA controller to synchronise the excitation from the transducer and reception from the lasers, as illustrated in Figure \ref{fig:Laser experimental setup}b. Data points were captured at a sampling frequency of 25 MHz but were upsampled to 1 GHz to achieve more accurate diffuse field correlation. Gel couplant was applied between the PA transducer and the sample surface.
 
As seen in Figure \ref{fig:Energy density curves}, the sensitivity of lasers to ultrasound is much lower compared to piezoelectric PA elements. This significantly impacts the reconstruction quality of the GF. Therefore, several modifications were implemented in the experiment to address this issue. Firstly, each set of diffuse signals underwent 2000 repetitions and averaging before storage. This extensive averaging aims to effectively reduce the unwanted random noise present in the signal. Secondly, each source consisted of four PA elements that were simultaneously excited to introduce more energy into the sample, thereby improving the SNR of the diffuse signals. Specifically, elements 1, 2, 3, and 4 were excited for the initial measurement, followed by elements 2, 3, 4, and 5 for the subsequent measurement, and so forth. Given the presence of 128 elements from 2 PA transducers, this slightly reduced the number of source averages from 128 to 122 ($N$ = 122). However, the simultaneous excitation of PA elements resulted in a more focussed beam of higher directivity, necessitating a longer duration before the wave energy reaches a diffused state. As a result of this phenomenon and the need for longer window length to improve the reconstruction accuracy as mentioned in Section \ref{Section: Parameter optimisation}, the reconstruction process required the use of diffuse signals at later times. Therefore, the diffuse fields were defined with a value of ${t_r}$ set at 0.2 ms and a much longer $T$ of 1800 \textmu s. In this case, there is no need to define a value larger than 1800 \textmu s because, the energy density curves showed that signals beyond this point in time are dominated by noise. Lastly, to account for the substantial amount of energy dissipation present within the 1800 \textmu s window length, 1B normalisation was applied to the diffuse signals before correlation to accelerate convergence towards GF reconstruction as discussed in Section \ref{Section: One-bit Normalisation}.

Both laser beams were aligned perpendicular to the sample surface to capture out-of-plane displacement and the laser inspection points were adjusted by rotating the aluminium sample with a rotation stage. This setup facilitated the reconstruction of the coherent response between the two points, enabling the determination of the arrival time of the longitudinal wave and, consequently, the measurement of the longitudinal velocity in five different directions as depicted in Figure \ref{fig:Laser experimental setup}a.

Each set of diffuse signals underwent the same procedure outlined in Section \ref{Section: PA experimental procedure} to calculate the velocity in each measurement direction and compared against the ground truth. Due to the curvature of the measurement surface and the localised nature of each laser measurement, achieving high-accuracy contact measurements at precisely the same spot for ground truth validation is challenging. Consequently, we opted to rely on velocity values from pulse-echo measurements and the material's isotropic characteristic to calibrate the laser results. The ground truth velocity was determined using the same 5 MHz PA transducer to capture pulse-echo signals in the out-of-plane direction of Figure \ref{fig:Laser experimental setup}a, at 128 different locations on the top flat surface of the sample. By calculating the TOF between two repeating back wall echoes and considering the sample's thickness, the velocity at each location was computed. The average of these calculated velocities was defined as the ground truth velocity. 

\begin{figure}[!t]
  \centering
  \includegraphics[width=0.9\textwidth]{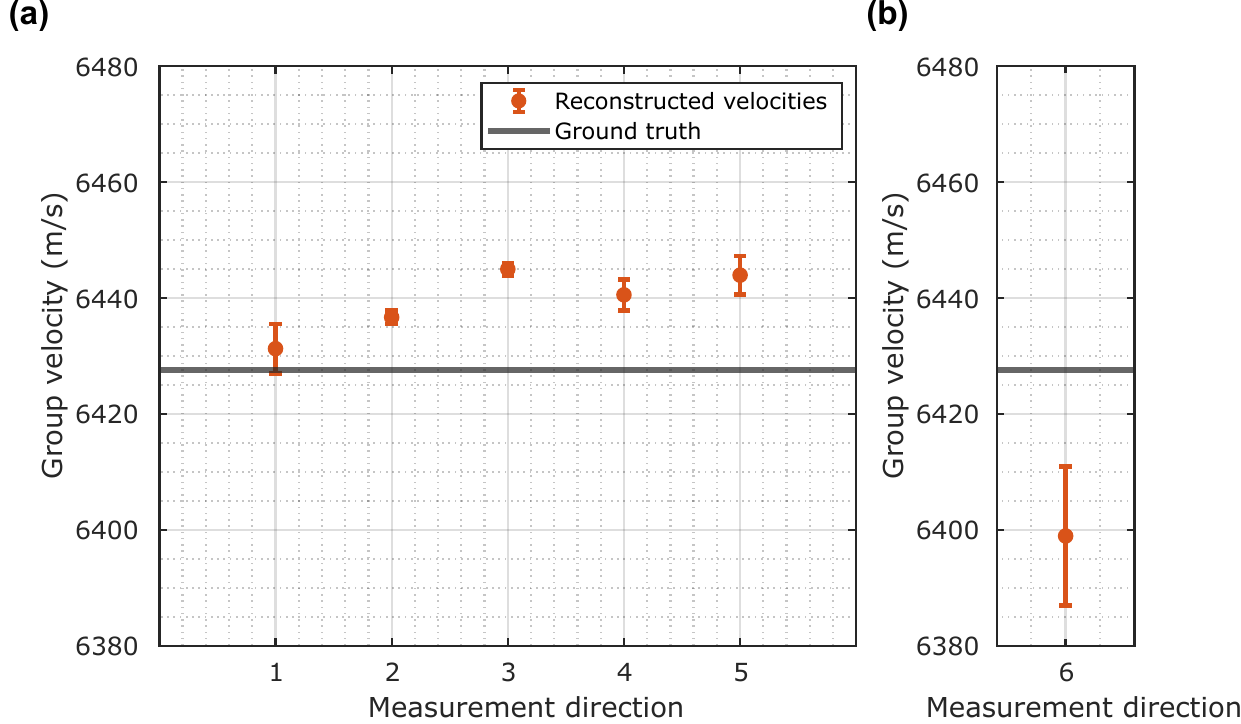}	
  \caption{Comparison of measured longitudinal group velocity from diffuse field reconstruction (orange data points) in \textbf{a.} five different normal incidence directions and \textbf{b.} one oblique direction with ground truth (solid grey line).}
  \label{fig:Laser velocity result}
\end{figure}

Three repetitions were conducted in each of the measurement directions from one to five. The mean values, along with standard deviations represented by error bars, were calculated and the outcomes are visually depicted in Figure \ref{fig:Laser velocity result}a. The average measured velocity using the GF reconstruction method in each direction deviates by less than 12 ms\textsuperscript{-1} from the ground truth velocity of 6427.6 ms\textsuperscript{-1}. The standard deviations hover around 4 ms\textsuperscript{-1}, likely due to fluctuations in experimental implementations. Additionally, all reconstructed velocities are higher than the ground truth, for which there are possible explanations. Although the sample block is isotropic, slight differences in velocity measured in different directions could exist. Therefore, the velocity measured in-plane (as shown in Figure \ref{fig:Laser experimental setup}a) using the lasers could differ from that measured out-of-plane, which was used to determine the ground truth. Alternatively, it could result from inaccurate positioning of laser inspection points such that the actual distance between the points are less than the diameter of the block. However, a difference of 12 ms\textsuperscript{-1} in velocity is equivalent to a difference in distance of 0.1 mm, which is challenging to improve further in practice. Lastly, the discrepancy could also stem from an unoptimised regularisation factor used during the deconvolution step, leading to errors in determining the TOF for velocity measurements. Nonetheless, this discrepancy is relatively small at about 12 ms\textsuperscript{-1}, indicating a significant similarity in measured velocities when using the GF reconstruction method.

A point to note is that all five measurement directions were normal incidence, which passes through the centre of the sample. This standardisation was done to define the distance between the two inspection points as a value that can be determined more accurately—specifically, the diameter of the sample—for velocity calculations. To validate the flexibility of lasers for wave speed measurements on complex geometries and in oblique directions, the coherent response between the top inspection point of direction one and the bottom inspection point of direction three, as seen in Figure \ref{fig:Laser experimental setup}a, was also reconstructed. Subsequently, we calculated the averaged velocity (based on three repetitions) in that direction, which is defined as direction six. The result for the oblique direction can be seen in Figure \ref{fig:Laser velocity result}b, which shows that the averaged measured velocity deviates from the ground truth by around 30 ms\textsuperscript{-1}. This deviation is acknowledged to be larger compared to the normal incidence measurement. However, we believe that this inaccuracy largely stems from the challenge of accurately measuring the distance between the inspection points, which could result in a corresponding measurement inaccuracy of about 0.3 mm. Overall, these results conclude that accurate velocity measurements in arbitrary directions of a material with curved geometry can be achieved using non-contact laser receivers. 

\section{Discussion} \label{Section: Discussion}
The results presented in the previous section highlight the effectiveness of the GF reconstruction method from diffuse fields for accurately estimating travel times between two points, thereby enabling precise velocity measurements. Our study also showcases the versatility of this methodology by demonstrating its applicability with different receivers, allowing for measurements on varied geometries. Notably, we achieved accurate measurements on samples with curved surfaces using laser receivers. The use of lasers in this methodology enables localised point-to-point wave speed measurements, a feat not achievable with conventional methods employing contact or immersion transducers. This capability is particularly advantageous for applications requiring fine spatial resolution and precision. The non-contact nature of lasers enhances the methodology's applicability to components with various complex geometries, such as additive manufactured parts, and allows for potential measurements in challenging environments like manufacturing processes. Furthermore, this method eliminates the use of more expensive high-power lasers for wave generation, thereby reducing equipment cost and omitting the associated safety requirements. Additionally, using laser receivers, which have a directivity pattern similar to that of piezoelectric elements \cite{Tu2024FiniteUltrasound}, to reconstruct the signals between two points eliminates the unfavourable directivity associated with laser ultrasonic generation in the thermo-optical region. Last but not least, when combined with the flexibility in source excitation offered by the GF reconstruction methodology, this approach unlocks opportunities for in-line material characterisation of manufactured components and a variety of other applications.

It is also important to acknowledge the limitations of the methodology, one of which is the increased measurement time associated with implementations using laser receivers. Due to the laser's lower sensitivity, it necessitates the averaging of multiple measurements to achieve accurate reconstruction. As a result, the measurement time can be lengthy, but this limitation could be alleviated by using lasers with better sensitivity. Additionally, reducing the wave source's central frequency or increasing the pulse voltage to introduce more energy into the sample could reduce the need for extensive averaging, thereby decreasing measurement times. Another limitation worth noting is the lower SNR of the reconstructed GF compared to directly using the coherent response. This factor introduces a trade-off to be weighed when deciding whether to adopt this methodology, taking into account its advantages alongside this limitation.

Nonetheless, this study suggests promising avenues for further research. One direction involves extending the methodology to reconstruct shear GF by measuring the in-plane displacement at laser inspection points, rather than the out-of-plane displacement. This adaptation would enable the measurement of shear velocity, potentially providing deeper insights into material characterisation due to the shear wave's higher sensitivity to material microstructure. Upon achieving this research goal, it could serve as an alternative and more accessible method for evaluating shear waves, considering that diffuse fields are dominated by shear energy, in a non-contact manner. Additionally, future efforts could focus also on expanding measurements to phase velocities. Given that texture measurement \cite{Lan2018DirectWaves} and many nondestructive techniques rely on phase velocity for determining elastic constants \cite{Rokhlin2011PhysicalComposites} this would require the inversion of group velocity measurements obtained from the GF reconstruction method.

Finally, ongoing research aims to investigate the impact of material diffusivity on the formation time of diffuse fields. Understanding how factors such as grain size and anisotropy affect diffusivity can provide valuable insights for measuring components made of diverse materials and geometries. This exploration will contribute to further enhancing the accuracy and applicability of the methodology in various industrial and research settings.

\section{Conclusion}

In this study, we explored the potential of employing GF reconstruction from ultrasonic diffuse fields for accurate velocity measurements on components with arbitrary geometries. The proposed application significantly increases the complexity compared to imaging applications, due to the sensitivity of arrival time to velocity measurements. Therefore, highly accurate reconstruction in terms of good waveform accuracy and minimal reconstruction noise must be achieved. Our approach involved establishing a set of necessary steps and techniques, which included the utilisation of spectral energy density curves to guide the optimisation of diffuse field parameters such as window length and time delay, the deconvolution of an unwanted source-dependent factor to achieve the correct waveform at the accurate arrival time and the implementation of 1B signal normalisation to expedite convergence to GF reconstruction within the same window length to account for energy decay in experiments. Subsequently, we demonstrated the capability of accurate velocity measurement using phased array receivers on a flat geometric sample and also in the more challenging implementation of laser receivers on a curved surface.

As such, this study establishes the methodology as an innovative tool for accurately measuring the velocity of materials with arbitrary geometries. Its successful demonstrations suggest its broader potential, extending material characterisation to materials of more complex geometries and a wider range of environments for different industrial applications, thus opening up opportunities for various industrial applications.

\section*{CRediT authorship contribution statement}
\noindent\textbf{Melody Png:} Conceptualisation, Data curation, Formal analysis, Investigation, Methodology, Project administration, Validation, Writing - Original draft, Writing - review and editing;

\noindent\textbf{Ming Huang:} Formal analysis, Funding acquisition, Investigation, Resources, Writing - review and editing;

\noindent\textbf{Marzieh Bahreman:} Resources, Writing - review and editing;

\noindent\textbf{Christopher M. Kube:} Resources, Writing - review and editing;

\noindent\textbf{Michael J.S. Lowe:} Formal analysis, Funding acquisition, Investigation, Project administration, Resources, Supervision, Writing - review and editing.

\noindent\textbf{Bo Lan:} Formal analysis, Funding acquisition, Investigation, Project administration, Resources, Supervision, Writing - review and editing.

\section*{Declaration of competing interest}
The authors declare that they have no known competing financial interests or personal relationships that could have appeared to influence the work reported in this paper.

\section*{Acknowledgement}
M.P. gratefully acknowledges the funding support from A*STAR Singapore under the National Science Scholarship. B.L. expresses gratitude for the support from the Imperial College Research Fellowship Scheme and the Non-Destructive Evaluation Group at Imperial College. B.L. and M.H. are thankful for the funding from EPSRC through Grant EP/W014769/1. C.K. acknowledges funding support from NSF through Grant No. 2225215.

\appendix
\renewcommand{\thefigure}{A\arabic{figure}}
\setcounter{figure}{0}
\renewcommand{\theequation}{A\arabic{equation}}
\setcounter{equation}{0}

\section*{Appendices}
\section{Definition of source-dependent term for deconvolution} \label{Appendix: Definition of source-dependent term for deconvolution}

 \begin{figure}[h!] 
    			\centering
      			\includegraphics[width=13cm, height=4.5cm]{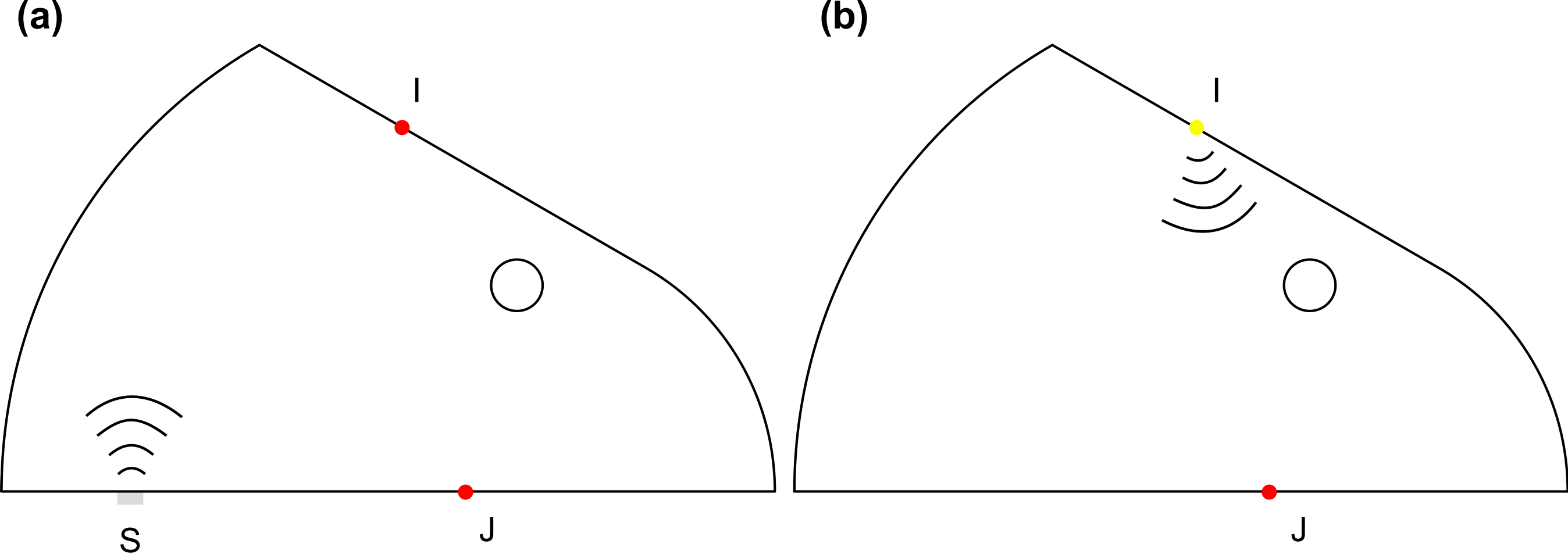}	
      			\caption{Schematic illustrating the setups used to describe the coherent response and the diffuse field correlation as a series of transfer functions. \textbf{a} illustrates the setup used to capture the diffuse fields at both location I and J from a source excitation at S to reconstruct the coherent response at J if there is a source at I. \textbf{b} illustrates the setup used to capture the coherent response at location J from a source excitation at location I.}
      			\label{fig: Understanding of source-dependent term via transfer functions}
 \end{figure}
 
Referring to Figure \ref{fig: Understanding of source-dependent term via transfer functions}a, let us represent the cross-correlation of diffuse fields received at points I and J from a source at S, denoted as $R_{J,I}(\omega)$, as a series of transfer functions. To do so, we first define the diffuse fields received at I and J as $D_{I,S}(\omega)$ and $D_{J,S}(\omega)$ respectively:
\begin{gather*}
   D_{I,S}(\omega) = W(\omega) \cdot S(\omega)  \cdot G_{I,S}(\omega) \cdot T_1(\omega) \\
   D_{J,S}(\omega) = W(\omega) \cdot S(\omega)  \cdot G_{J,S}(\omega) \cdot T_2(\omega)
\end{gather*} 

Thereafter, we can define the diffuse field correlation $R_{J,I}(\omega)$ as:
\begin{equation}\label{eq:A1}
\begin{gathered}
  R_{J,I}(\omega) = D_{I,S}(-\omega) \cdot D_{J,S}(\omega) \\ = W(-\omega) \cdot S(-\omega)  \cdot G_{I,S}(-\omega) \cdot T_1(-\omega) \cdot W(\omega) \cdot S(\omega)  \cdot G_{J,S}(\omega) \cdot T_2(\omega)
\end{gathered}
\end{equation}

where $W(\omega)$ represents the excitation waveform, $T_1(\omega)$ is the transfer function of the receiver at location I, $T_2(\omega)$ is the transfer function of the receiver at location J, and $S(\omega)$ is the transfer function of the source at location S. $G_{I,S}(\omega)$ and $G_{J,S}(\omega)$ denote the Green's Function between locations I and S, and between locations J and S, respectively.

 Figure \ref{fig: Understanding of source-dependent term via transfer functions}b illustrates the configuration used to capture the coherent response at point J from a source at point I, denoted as $C_{J,I}(\omega)$, which is the signal we aim to reconstruct from the setup depicted in Figure \ref{fig: Understanding of source-dependent term via transfer functions}a. The coherent response can be expressed as:

 \begin{equation}\label{eq:A2}
  		C_{J,I}(\omega) = W(\omega) \cdot T_1(\omega) \cdot G_{J,I}(\omega) \cdot T_2(\omega)
 \end{equation}

 where $W(\omega)$ is the excitation waveform used by the source to generate the diffuse field in Equation \ref{eq:A1}, $T_1(\omega)$ is the transfer function of the source at location I, $T_2(\omega)$ is the transfer of the receiver at J, and $G_{J,I}(\omega)$ is the GF between locations I and J. 
 
 With reference to Derode's paper \cite{Derode2003RecoveringL}: 
 
 \begin{equation}\label{eq:A3}
  		G_{I,S}(-\omega) \cdot G_{J,S}(\omega) = G_{J,I}(\omega) \cdot G_{S,S}(-\omega)
 \end{equation}
 where $G_{S,S}(-\omega)$ is the time-reversed GF at location S.
 
 Substituting Equation \ref{eq:A3} into the Equation \ref{eq:A1}, the equation that describes the cross-correlation of diffuse fields can be re-written as:

 \begin{equation}\label{eq:A4}
  		R_{J,I}(\omega) = W(-\omega) \cdot T_1(-\omega) \cdot S(-\omega) \cdot W(\omega) \cdot S(\omega) \cdot T_2(\omega) \cdot G_{J,I}(\omega) \cdot G_{S,S}(-\omega)
 \end{equation}
 
 By comparing Equation \ref{eq:A4} to Equation \ref{eq:A2}, we can identify the terms that need to be removed for the diffuse field correlation to yield the coherent response, which can be defined by $F(-\omega)$:
 
 \begin{equation}\label{eq:A5}
  	  F(-\omega) = \frac{R_{J,I}(\omega)}{C_{J,I}(\omega)} = W(-\omega) \cdot S(-\omega) \cdot S(\omega) \cdot G_{S,S}(-\omega)  
 \end{equation}
 Here, the time-reversed receiver function $T_1(-\omega)$ is assumed to be equivalent to the source function $T_1(\omega)$.


\end{document}